\documentclass[letterpaper]{article} 
\usepackage[]{aaai2026}  
\usepackage{times}  
\usepackage{helvet}  
\usepackage{courier}  
\usepackage[hyphens]{url}  
\usepackage{graphicx} 
\usepackage{natbib}  
\usepackage{caption} 
\usepackage{enumitem}

\title{Data Annotation as Measurement}
\author {
    Emma Harvey\textsuperscript{\rm 1, 2},
    Allison Koenecke\textsuperscript{\rm 1, 2},
    René F. Kizilcec\textsuperscript{\rm 2}
}
\affiliations {
    \textsuperscript{\rm 1}Cornell Tech, 
    \textsuperscript{\rm 2}Cornell University\\
    evh29@cornell.edu, koenecke@cornell.edu, kizilcec@cornell.edu
}

\usepackage{subcaption}
\usepackage{rotating}
\usepackage{multirow}
\usepackage{colortbl}
\usepackage{tabularx}
\usepackage{xcolor}
\usepackage{array}
\usepackage{booktabs}
\usepackage{amssymb}

\newcolumntype{M}[1]{>{\raggedright\arraybackslash}m{#1}}

\usepackage{siunitx}

\begin{document}

\maketitle

\begin{abstract}
Modern AI systems depend on annotated data, but annotation is rarely treated as the act of measurement that it is. Instead, annotation quality is commonly reduced to agreement: if multiple annotators assign the same annotation to a data instance, the annotations are taken to be high-quality. Yet agreement does not establish whether annotations validly capture the underlying concept they are meant to represent. In this paper, we argue that data annotation should be understood as a measurement problem. Like other forms of measurement, annotation requires defining a concept, operationalizing it through an instrument, applying that instrument, and evaluating the reliability and validity of the resulting measurements. Drawing on a literature review of annotation quality research (N=132) and semi-structured interviews with annotation team members (N=10), we develop a framework for diagnosing and correcting annotation issues. First, we map key decision points across annotation processes\textemdash including task design, annotator management, quality assessment, quality improvement, and adjudication\textemdash that shape annotation outcomes. Second, we identify five distinct sources of annotation issues: error, ambiguity, impossibility, subjectivity, and annotator identity. Annotation problems that appear similar at the level of outcomes often require different process-level interventions based on their sources. Finally, we translate measurement theory into practical guidance for annotation teams, showing how assessments of reliability and validity can move beyond agreement alone. By reframing annotation as measurement, we offer a conceptual foundation for improving the quality of annotated data used in AI research and practice.\looseness=-1
\end{abstract}

\section{Introduction}\label{sec:introduction}
Modern AI systems are built and evaluated using data. This data is often produced through \textit{data annotation}, the process of augmenting raw data so that it can be used for some downstream task. In this work, we argue that annotation should be thought of as a form of measurement in which teams systematize and operationalize concepts and apply those operationalizations to data instances. To date, various best practices for data annotation have been proposed---but, despite the centrality of annotation to AI development, few have been widely adopted~\cite{klie_analyzing_2024}. Perhaps as a result, much annotated data is of shockingly poor quality. Researchers have found that label errors in widely-used evaluation benchmarks obfuscate measures of model performance, and that datasets produced by crowdworkers contain issues including logical fallacies and failures of basic quality control and consistency~\cite{blodgett_stereotyping_2021, harvey_framework_2025, northcutt_pervasive_2021}. \citet{harvey-etal-2025-understanding} recently found that these issues are so pervasive that researchers and practitioners often struggle to use any annotated data at all.\looseness=-1

To date, researchers have primarily identified these issues by calculating agreement, either between annotations and ``ground truth,'' or between multiple annotators. While this approach can reveal issues, we argue that it is insufficient for correcting them for two primary reasons. First, correcting issues at scale requires changing annotation processes, not only annotation outcomes.\footnote{Using ``ground truth'' to identify issues is only possible if ground truth is available, and if ground truth were available for each data instance, annotation would not be needed. Correcting issues even when ground truth is not available thus requires changing an aspect of the annotation process and re-annotating.\looseness=-1} But annotation processes, like data work in general, are under-specified and under-documented, leaving key decision points that lead to annotation issues invisible: treated as defaults, not documented, and not interrogated~\cite{paullada_data_2021, sambasivan_everyone_2021, scheuerman_datasets_2021}. Second, annotation teams often rely on agreement among annotators as their primary quality signal~\cite{klie_analyzing_2024}. Agreement is simple to calculate and useful as one indicator of reliability, but treating it as evidence of validity is a mistake: annotators may agree with one another while systematically misinterpreting the concept, applying the same flawed assumptions, or following ambiguous instructions in the same way~\citep{catanzariti_taming_2025, jacobs_measurement_2021, zhao_assumptions_2013}.\looseness=-1

In this work, we present the results of a qualitative study intended to support the correction of data annotation issues. Informed by a semi-systematic literature review of research on annotation quality (N=132) and semi-structured interviews of researchers on annotation teams (N=10),\footnote{\textit{Annotation teams} include \textit{annotators} and \textit{annotation managers}, who design and manage annotation tasks.} we make two primary contributions. First, \textbf{we make annotation processes visible.} We map key decision points throughout the annotation process that can lead to issues, showing where annotation processes can be changed to improve annotation outcomes. As part of this, we identify five distinct sources of annotation issues: error, ambiguity, impossibility, subjectivity, and annotator identity. These issues can all manifest as a misalignment between annotations and ``ground truth,'' but they arise from different parts of the annotation process and therefore require different interventions.\looseness=-1

Second, \textbf{we argue that researchers and practitioners should treat annotation as a measurement problem}. In so doing, we build on prior work that has called for measurement theory to be applied to sociotechnical research~\cite{jacobs_measurement_2021, wallach_position_2025} and provide a framework for grounding calls to move beyond agreement when evaluating data annotation~\cite{thomas2026modernizinggroundtruthshifts}. Drawing on measurement theory, we show how annotation teams can evaluate not only whether annotators agree, but whether annotations are reliable and valid. These contributions shift annotation quality from an outcome-level problem that asks whether labels agree with one another or with a nominal ground truth to a measurement problem that asks whether annotation processes produce reliable and valid representations of the concepts they purport to capture.\looseness=-1

\section{Background}\label{sec:background}
Annotation, a component of data development~\cite{paullada_data_2021}, is the process of augmenting data (text, image, audio, video, etc.) so that it can be used for some downstream task. There are three types of tasks that are commonly viewed as annotation~\cite{shmueli_beyond_2021}: 
\begin{itemize}[leftmargin=*]
    \item \textit{Labeling}: identification (e.g., drawing a bounding box around people in an image) and categorization (e.g., determining the species of a plant in an image)
    \item \textit{Evaluation}: judging a data instance, according to a rubric (e.g., scoring an essay) or to annotator preference (e.g., selecting which AI-generated response to a prompt is better)
    \item \textit{Production}: producing new data (e.g., captioning a video)
\end{itemize}
Annotation tasks range from more objective (e.g., identifying the speaker in a video) to extremely subjective (e.g., determining the intent of the speaker).\looseness=-1

\paragraph{Issues in Data Annotation.}
``Data work''---the time and effort that goes into data development---has long been undervalued in ML research and practice~\cite{sambasivan_everyone_2021,gray2019ghost}, and as a result, the field has coalesced around data development practices that prioritize efficiency and scale over quality~\cite{paullada_data_2021}. Data annotation is no exception~\cite{scheuerman_datasets_2021}. At the same time, it is hard to fully account for the issues present in data annotation because, as \citet{klie_analyzing_2024} point out, many papers do not describe their annotation process in any capacity---despite the wide variety of documentation practices proposed by researchers~\cite{hutchinson_towards_2021, gebru_datasheets_2021, hallinan_dataset_2020, fabris_tackling_2022, diaz_crowdworksheets_2022, bender_data_2018}. Therefore, issues in data annotation are typically identified by looking at the outcomes of previously conducted annotation processes. For example, researchers have identified widespread errors in labeled data (e.g., where images are misidentified)~\cite{northcutt_pervasive_2021, klie_annotation_2023} as well as in produced data (e.g., where crowdsourced text deviates from annotation instructions)~\cite{blodgett_stereotyping_2021, harvey_framework_2025}. \citet{harvey-etal-2025-understanding} found that practitioners regularly identify label errors in publicly available annotated data, although they generally do not document or publish those findings. In other cases, issues are identified by looking at the performance of models trained or evaluated on datasets. For example, researchers have found that image classification models trained on widely-used datasets like ImageNet can fail to generalize to novel images~\cite{pmlr-v97-recht19a}, at least in part because label categories are too loosely defined~\cite{torralba_unbiased_2011}.\looseness=-1

\textit{Ground Truth.} Prior work has often identified issues as occurring when annotations are misaligned with a ``ground truth.'' It is important to note that \textit{ground truth} itself is nebulous: it refers to the true value of the underlying concept that an annotation seeks to measure, and for any given data instance it may be unknown or contested---and the assumption that a single ground truth even exists may be flawed~\cite{Sloane2026}.\footnote{It is possible, however, to identify misalignments between an annotation and a ground truth even when ground truth is not known. For example, it is possible to determine that an annotation identifying a photo of a tree as a strawberry is incorrect even without knowing the ground truth (i.e., what species of plant the tree is).\looseness=-1} Thus, researchers often use proxies to identify misalignment between annotations and ground truth. For example, researchers may produce ``gold labels'' (which are often expert annotations) for a subset of data instances and treat gold labels as ground truth~\cite{snow_cheap_2008}.\looseness=-1

\textit{Agreement.} 
Although work reviewing annotation outcomes often compares annotations to ``ground truths,'' annotation processes usually focus on maximizing \textit{agreement} or \textit{inter-rater reliability} (IRR). Agreement refers to the extent to which multiple annotations of the same data instance produced by different annotators agree with one another~\cite{krippendorff_alpha, cohen_kappa, mcdonald_reliability_2019}. It is appealing because it is easy to calculate: as long as multiple annotations are collected for each data instance, agreement can be calculated for any labeling task whether or not ground truth is available.\footnote{One limitation of agreement, however, is that it is not necessarily meaningful for preference-based evaluation or production tasks.} In many cases, annotation teams treat agreement as a proxy for ground truth. Annotation managers aggregate annotations (e.g., by keeping the majority annotation for a data instance) and then treat the aggregated label as a ``silver label''~\cite{uma_learning_2021}. However, prior work has repeatedly identified limitations of and assumptions inherent to this process, including that highly reliable annotations do not necessarily align with ground truth~\cite{zhao_assumptions_2013, reidsma_reliability_2008, bayerl_what_2011, thomas_beyond_2025, arhin_ground-truth_2021, Ali_Zhao_Koenecke_Papakyriakopoulos_2026}.\looseness=-1

\subsection{Measurement Theory}\label{subsec:background:measurement}
Measurement theory from the social sciences describes the process by which underlying concepts---i.e., ground truths, which may be unobservable, theoretical, or contested---are captured in instance-level measurements~\cite{adcock_measurement_2001, wallach_position_2025}. Measurement, as outlined by \citet{adcock_measurement_2001} is a multi-stage process. The first stage, \textit{systematization}, involves giving an explicit definition to a concept of interest. The next stage, \textit{operationalization}, involves developing processes or instruments through which to measure the systematized concept. Measurements are produced for specific data instances in the \textit{application} stage. Finally, \citet{wallach_position_2025} argue that measurement processes should include an \textit{interrogation} stage, in which the reliability and validity of measurements are assessed. \textit{Reliability} asks whether a measurement can be repeated, while \textit{validity} asks whether a measurement is correct~\cite{jacobs_measurement_2021}. Researchers have identified multiple aspects of reliability and validity that can be interrogated~\cite{cronbach_construct_1955, messick_validity_1987, messick_validity_1995, cook_current_2006, jackman_measurement_2008, jacobs_measurement_2021}. We use these aspects of reliability and validity to suggest opportunities to improve data annotation in \S\ref{sec:results:measurement}.\looseness=-1

\paragraph{Measurement and Data Annotation.} In recent years, researchers have increasingly argued that various aspects of sociotechnical research should be viewed through the lens of measurement theory~\cite{jacobs_measurement_2021, wallach_position_2025}. Building on these arguments, a large body of work has used reliability and validity to interrogate data~\cite{blodgett_stereotyping_2021, eriksson_can_2025, gehrmann_repairing_2023, gosciak_scrutinizing_2026, harvey-etal-2025-understanding, thomas_beyond_2025, thomas2026modernizinggroundtruthshifts, van_der_wal_undesirable_2024, xiao_evaluating_2023, zhao_position_2024, zhou_deconstructing_2022}. For example, both \citet{blodgett_stereotyping_2021} and \citet{eriksson_can_2025} use interrogation to identify major validity issues in widely-used AI and ML benchmarks (including those based on labeled or produced data), and call for increased consideration of validity during dataset development. By building on this broad base of prior work to explicitly treat data annotation as a measurement problem, we offer a framework to ground existing and future work on evaluating data annotation.\looseness=-1
\section{Methods}\label{sec:methods}
Our qualitative methods followed an iterative and highly intertwined approach. First, we conducted a semi-systematic literature review of papers related to annotation quality (\S\ref{subsec:methods:litreview}). We found two categories of papers: those that focused on annotation \textit{processes} and those that focused on annotation \textit{outcomes}. Within the papers that focused on processes, we found that papers tended to either propose ``best practices'' or survey ``current practices'' (at least to the extent that current practices are documented in published work). However, we struggled to find papers that described \textit{how} annotation teams apply current or best practices---i.e., what decisions annotation teams make throughout the annotation process that affect the quality of annotations. Thus, we supplemented our literature review with a small-scale semi-structured interview study of researchers on annotation teams in order to surface those accounts (\S\ref{subsec:methods:interview}). 

\subsection{Literature Review}\label{subsec:methods:litreview}
We conducted a semi-systematic literature review of academic work related to annotation quality.\footnote{A semi-systematic literature review is a research method that is designed for tasks like providing an overview of research that spans diverse disciplines (like data annotation). While semi-systematic literature reviews do not necessarily require strictly systematic search and research strategies (e.g., following PRISMA guidelines), they nevertheless require that strategies are transparently presented and well justified~\cite{SNYDER2019333}.\looseness=-1} We conducted a keyword search of the ACM, ACL, IEEE, and NeurIPS paper databases for terms related to annotation (\texttt{annotat* | label* | coding}) and quality (\texttt{subjectiv* | objectiv* | ambigu* | eval* | valid* | reliab* | *agree* | bias* | ground truth | error*}), requiring that keywords representing both annotation and quality were present in the title or abstract of the work. We selected relevant works by manually reviewing and filtering papers, first by title and then by abstract. Inclusion criteria for relevance were: 
\begin{itemize}
    \item[(a)] Being about annotation specifically (as opposed to, e.g., crowdwork in general); and
    \item[(b)] Evaluating, critiquing, or improving annotation outcomes (as opposed to, e.g., producing annotations); or 
    \item[(c)] Proposing or documenting annotation practices.
\end{itemize}
\noindent We supplemented this using a natural language search for \texttt{``barriers or best practices in data annotation''} using AI2 Asta.\footnote{\url{https://asta.allen.ai/}} Finally, to minimize selection bias, we reviewed the citations of papers surfaced by our search and added work that met the inclusion criteria for title and abstract relevance to our pool. Overall, we reviewed 132 publications spanning 1972 to 2026.\looseness=-1

We categorized papers as focused on annotation processes or annotation outcomes. Papers focused on annotation processes either proposed ``best practices'' or surveyed ``current practices,'' while papers focused on annotation outcomes evaluated the quality of existing annotated data. We synthesized process-focused papers to develop a high-level understanding of the stages of the annotation process, which we used to scaffold our semi-structured interviews. We categorized outcome-focused papers according to their evaluation approach, attempting to map evaluation approaches to one or more of the reliability and validity aspects described in \S\ref{subsec:background:measurement}.\looseness=-1

\subsection{Interview Study}\label{subsec:methods:interview}

\begin{table*}[ht]
\begin{small}
\centering
\begin{tabular}{c|
c!{\color{lightgray}\vrule}
c|
c!{\color{lightgray}\vrule}
c!{\color{lightgray}\vrule}
c|
c!{\color{lightgray}\vrule}
c!{\color{lightgray}\vrule}
c|
c!{\color{lightgray}\vrule}
c!{\color{lightgray}\vrule}
c|
c!{\color{lightgray}\vrule}
c!{\color{lightgray}\vrule}
c|
c!{\color{lightgray}\vrule}
c|}
\multicolumn{1}{c}{}
& \multicolumn{2}{c}{\textbf{Role}} &
\multicolumn{3}{c}{\textbf{Medium}} &
\multicolumn{3}{c}{\textbf{Domain}} &
\multicolumn{3}{c}{\textbf{\# Projects}} &
\multicolumn{3}{c}{\textbf{Race}} &
\multicolumn{2}{c}{\textbf{Gender}} \\
\arrayrulecolor{black}\cline{2-17}
& \begin{sideways}\textit{Annotation Manager\textcolor{white}{..}}\end{sideways} & \begin{sideways}\textit{Annotator}\end{sideways} &
 \begin{sideways}\textit{Text}\end{sideways} &
 \begin{sideways}\textit{Image}\end{sideways} &
 \begin{sideways}\textit{Video}\end{sideways} &
 \begin{sideways}\textit{General-Purpose ML}\end{sideways} &
 \begin{sideways}\textit{Education}\end{sideways} &
 \begin{sideways}\textit{Law}\end{sideways} &
 \begin{sideways}\textit{1 to 5}\end{sideways} & \begin{sideways}\textit{6 to 10}\end{sideways} & \begin{sideways}\textit{more than 10}\end{sideways} & \begin{sideways}\textit{Asian}\end{sideways} & \begin{sideways}\textit{Black}\end{sideways} & \begin{sideways}\textit{White}\end{sideways} & \begin{sideways}\textit{Male}\end{sideways} & \begin{sideways}\textit{Female}\end{sideways} \\
\arrayrulecolor{black}\hline
\textit{P1} & \checkmark & \checkmark &
\checkmark & & \checkmark & 
& \checkmark &&
& \checkmark &&
&& \checkmark &
 \checkmark & \\
\arrayrulecolor{lightgray}\hline
\textit{P2} & \checkmark & \checkmark & 
\checkmark &&&
&& \checkmark &
\checkmark &&&
&& \checkmark &
 \checkmark & \\
\arrayrulecolor{lightgray}\hline
\textit{P3} & \checkmark & \checkmark & 
\checkmark &&& 
&& \checkmark &
\checkmark &&&
&& \checkmark &
& \checkmark \\
\arrayrulecolor{lightgray}\hline
\textit{P4} & \checkmark & \checkmark & 
\checkmark & \checkmark &&
 \checkmark &&&
\checkmark &&&
\checkmark &&&
& \checkmark  \\
\arrayrulecolor{lightgray}\hline
\textit{P5} & 
\checkmark & \checkmark & 
\checkmark &&&
&& \checkmark &
\checkmark &&&
&& \checkmark &
 \checkmark & \\
\arrayrulecolor{lightgray}\hline
\textit{P6} & \checkmark & \checkmark & 
\checkmark &&&
&& \checkmark &
\checkmark &&&
&& \checkmark &
&  \checkmark \\
\arrayrulecolor{lightgray}\hline
\textit{P7} & \checkmark & \checkmark & 
\checkmark &&&
& \checkmark &&
\checkmark &&&
& \checkmark & \checkmark &
 \checkmark & \\
\arrayrulecolor{lightgray}\hline
\textit{P8} & & \checkmark & 
\checkmark &&&
&& \checkmark &
\checkmark &&&
\checkmark &&&
&  \checkmark \\
\arrayrulecolor{lightgray}\hline
\textit{P9} & \checkmark & \checkmark & 
\checkmark & \checkmark & & 
 \checkmark &&&
&& \checkmark &
\checkmark &&&
 \checkmark & \\
\arrayrulecolor{lightgray}\hline
\textit{P10} & \checkmark & \checkmark &
\checkmark & \checkmark &&
& \checkmark &&
&& \checkmark &
&& \checkmark &
&  \checkmark \\
\arrayrulecolor{black}\hline
\textbf{Totals} & 9 & 10 & 10 & 3 & 1 & 2 & 3 & 5 & 7 & 1 & 2 & 3 & 1 & 7 & 5 & 5 \\ 
\arrayrulecolor{black}\hline
\end{tabular}
\caption{Characteristics of interview participants, including demographics and prior experience with annotation projects. Note that participants could report multiple roles, data mediums, domains, and racial identities.\looseness=-1}
\label{tab:participants}
\end{small}
\end{table*}

Between November and December of 2025, we conducted 10 semi-structured interviews with individuals who have been involved in data annotation projects. We recruited participants through purposive and snowball sampling in order to recruit individuals with varied experience with and opinions on data annotation. Interviews were 60 minutes long and were conducted and recorded via Zoom. Participants provided informed consent prior to the start of the interview, and all participants were compensated for their time with a \$50 gift card. The study was reviewed by an Institutional Review Board (IRB).\looseness=-1

\subsubsection*{Participants.} 
All participants have taken on annotator roles in prior projects, and all but one have also taken on annotation manager roles. Participants work across a variety of domains, including education, law, and general-purpose ML. All participants hold research roles, although several described working on annotation projects within industry labs or for non-academic reasons (e.g., for the purposes of mounting a legal appeal). All participants are based in North America. Additional background information about interview participants is provided in Table~\ref{tab:participants}.\looseness=-1 

\subsubsection*{Interviews.} 
To start, participants were asked to briefly list the data annotation projects that they had been involved with. They were then asked to describe each project, including the annotation goal, data medium (e.g., text, image, video), scale (e.g., data size, project timeline), and annotator population (e.g., experts, trained annotators, crowdworkers, AI). Participants were asked to describe their annotation processes in detail, focusing on key decision points and challenges that they faced throughout. We initially asked open-ended questions (e.g., ``can you describe the data annotation process on your project?'') and asked more specific follow-ups as needed (e.g., ``what tools or platforms did you use to collect annotations?''). For all decision points and challenges mentioned, participants were asked what action they took and what they thought the effects of that action on their resulting annotations were. The full semi-structured interview guide is available in Appendix \ref{app:interviewguide}.\looseness=-1

\subsubsection*{Thematic Analysis.} We then conducted a reflexive thematic analysis on the resulting interview transcripts, following an inductive-deductive coding approach~\cite{braun_using_2006, braun_reflecting_2019}. The first author, in discussion with the other authors, developed an initial set of deductive codes that focused on key decision points related to misalignment between annotations and a ``ground truth,'' including how participants identified misalignments and what they did in their annotation processes to address them. The first author then manually reviewed the transcripts to identify the presence of those codes. While doing so, the first author identified additional key decision points raised by participants and inductively added them to the set of codes. Finally, the first author recoded all transcripts with the full set of codes.\looseness=-1

\subsubsection*{Positionality.}
Our approach to thematic analysis is shaped by our own positionality. We are a team of academic researchers with backgrounds in computer science, data science, statistics, and philosophy. Before their career in academia, the first author worked in industry as a data scientist, and, as a result, personally values data work and is liable to notice and question when it is implicitly undervalued. At the same time, although all members of the research team have served as annotation managers or personally annotated data for various research projects, none have ever worked as crowdworkers or as full-time data annotators. Furthermore, because we are based in the US and English is the only language shared across all members of the research team, our interviews were conducted in English and our analysis is US-centric.\looseness=-1

\paragraph{Limitations.}
We acknowledge several limitations of our approach. In particular, our interview study relied on a small (though in line with typical sample sizes for similar research; see \citet{caine}) and non-representative participant pool. We believe our interview study is nevertheless valuable as a supplement to our literature review, which lacked narrative accounts of decision points and associated challenges arising during annotation projects. Although our participant pool is not representative of all annotators or annotation managers, it is not intended to be: we used targeted recruiting to reach out to individuals who we knew had varied experiences with and opinions on annotation in order to capture a breadth of perspectives.\looseness=-1
\section{Making Annotation Processes Visible}\label{sec:results:processes}
In this section, we focus on annotation processes. Drawing on our literature review, we provide a brief overview of the high-level annotation process (\S\ref{subsec:results:processes:process}). Then, drawing primarily on our interview study, we provide a more detailed map of key decision points that occur throughout that process and that can lead to annotation issues (\S\ref{subsec:results:processes:decisionpoints}). Finally, drawing on both studies, we identify distinct sources of annotation issues stemming from different decision points (\S\ref{subsec:results:sources}).\looseness=-1

\subsection{High-Level Annotation Process}\label{subsec:results:processes:process}

\begin{figure}
\centering
\includegraphics[width=0.46\textwidth]{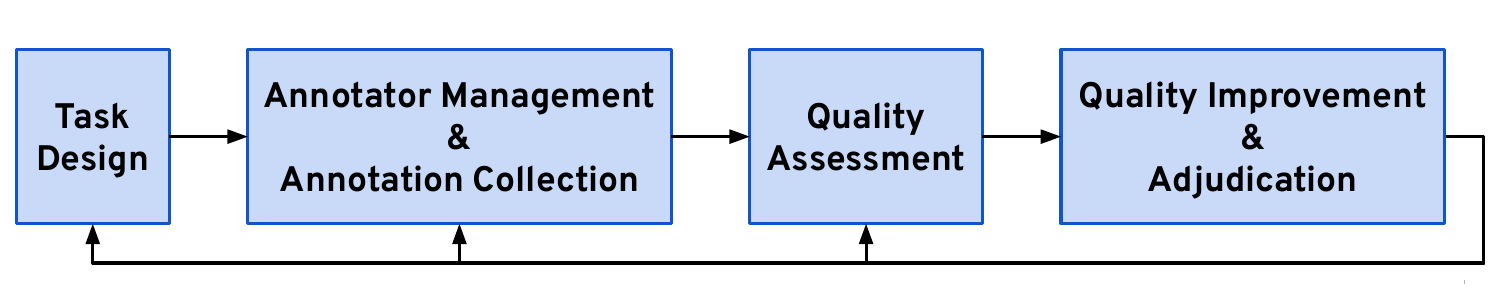}
\caption{\small The high-level data annotation process consists of iterative stages of Task Design, Annotator Management \& Annotation Collection, Quality Assessment, and Quality Improvement \& Adjudication.\looseness=-1}\label{fig:annotationprocess}
\end{figure}

There are multiple disciplines across which data annotation is conducted: while some projects draw on principles of qualitative coding from the social sciences, others view annotation as a more straightforward crowdsourcing challenge---and as AI-based annotation becomes widespread, annotation is increasingly seen as a prediction task~\cite{callison-burch_creating_2010, he_if_2024, macqueen_codebook_1998}. Different disciplines follow slightly different annotation processes, but all generally include the four stages of Task Design, Annotator Management \& Annotation Collection, Quality Assessment, and Quality Improvement \& Adjudication, as shown in Figure~\ref{fig:annotationprocess}~\cite{beck_quality_2023, daniel_quality_2019, fereday_demonstrating_2006, klie_analyzing_2024, macqueen_codebook_1998, sabou_corpus_2014, tornberg_best_2024}. 

\textit{Task Design} includes selecting data, writing a codebook or annotation guidelines, and planning what the annotation process will look like. \textit{Annotator Management \& Annotation Collection} includes selecting an annotator population, training annotators, and collecting annotations. \textit{Quality Assessment} includes determining whether already-collected annotations are of sufficiently high quality. If quality is not sufficiently high, \textit{Quality Improvement} involves changing either annotation processes or annotation outcomes to improve annotation quality; otherwise, \textit{Adjudication} involves selecting the final set of annotations from all those collected.\looseness=-1

\subsection{Key Decision Points in Annotation Processes}\label{subsec:results:processes:decisionpoints}

\begin{table*}
\small
\centering
\begin{tabular}{>{\hspace{0pt}}m{0.14\linewidth}>{\hspace{0pt}}m{0.12\linewidth}>{\hspace{0pt}}m{0.6\linewidth}} 
\toprule
\textbf{Stage} & \textbf{Components} & \textbf{Key Decision Points} \\
\midrule
\multirow{8}{\linewidth}{\textbf{Task Design}} & 
Data & 
Is data representative of the underlying concept?\par{}
How should data be preprocessed?\\
\cmidrule{2-3}
& Schema & 
What is the modality of annotation?\par{}
What is the unit of annotation?\par{}
Should measures of certainty be collected? How?\\
\cmidrule{2-3}
& Guidelines & 
How to handle the comprehensiveness/readability tradeoff?\\
\cmidrule{2-3}
& Process & 
Through what interface should annotations be collected?\par{}
What is the optimal level of coordination among annotators?\\
\midrule
\multirow{5}{\linewidth}{\textbf{Annotator Management \& Annotation Collection}} &
Selection & 
What should the annotator population be?\par{}
Should screening or qualification testing be applied?\\
\cmidrule{2-3}
& Training \&\par{}Feedback & 
How should annotators be trained?\par{}
Should annotator feedback be provided/solicited?\\
\cmidrule{2-3}
& Incentives & What is the optimal payment structure? \\
\midrule
\textbf{Quality}\par{}\textbf{Assessment} & Metric Selection & Should agreement be used? If so, which agreement metric should be used?\par{}
Should gold labels be created? If so, how?\\
\midrule
\multirow{4}{\linewidth}{\textbf{Quality}\par{}\textbf{Improvement} \textbf{\&}\par{}\textbf{Adjudication}} & Improvement & 
Should new annotations be collected or should existing annotations be filtered?\par{}
If new annotations are collected, what changes should be made to the process?\par{}
If existing annotations should be filtered, how?\\
\cmidrule{2-3}
& Adjudication & 
Should all annotations be retained, or should annotations be aggregated?\par{}
If annotations are to be aggregated, how?\par{}
How should certainty be expressed?\\
\bottomrule
\end{tabular}
\caption{Key decision points affecting the quality of annotations at each stage of the annotation process, identified through our literature review and interview study.}\label{tab:decisionpoints}
\end{table*}

Each stage of the annotation process has associated key decision points. These decision points, many of which are undocumented in published annotation studies and even treated as ``defaults'' by annotation teams, have the potential to affect the quality of annotation outcomes. We summarize key decision points in Table~\ref{tab:decisionpoints} and discuss them in detail below.

\paragraph{Task Design.}
Before annotation can begin, data must be collected and pre-processed so that it is well-formatted for annotation. Key decision points in this stage include determining whether data is representative of the underlying concept that annotation seeks to capture and determining how data should be preprocessed. Interview participants who used ``gold labels'' reported that it was particularly important to determine whether those labels were representative of their underlying concept, as non-representative gold labels will systematically fail at catching annotation issues (P2, P5). Interview participants also reported struggling to determine the best way to preprocess data without introducing ``\textit{compounding}'' (P1) or ``\textit{cascading}'' (P10) issues (P1--P3, P10). For example, P2 described an annotation task in which crowdworkers evaluated the clarity of website terms of service. To set up the task, he scraped terms of service from sites and had to decide whether to strip HTML elements from the scraped pages (which risked dropping relevant information like bolding, font size, and font color) or to leave them in (which risked reducing readability of the text). Although prior work has similarly found that annotation quality is affected by data collection and preprocessing procedures, such decisions are rarely discussed in published annotation studies~\cite{grondin-verdon_qualitative_2024, hays_simplistic_2023}.\looseness=-1

A decision that was particularly challenging for many participants was determining the unit of annotation (P1--P3, P6, P8, P10). For example, if text data is annotated, should it be annotated sentence-by-sentence or paragraph-by-paragraph? P6, who was annotating trial transcripts for aspects of gender stereotyping (e.g., hypersexualizing women), described the tradeoffs inherent to this decision:\looseness=-1
\begin{quote}
``\textit{When you have a trial transcript\ldots you have a lot of interruptions\ldots So this is a little bit more difficult to code by sentence\ldots it was very difficult for researchers to [know whether they should] code [an exchange] three different times, or [just] once. And we in the end decided that it would increase inter-rater reliability if we expanded the unit of text\ldots That creates a problem, because if, quantitatively, you now want to say, ‘how many times was there a reference to her having sex, or wanting to have sex?’ we've only coded it once, instead of coding it three times. And what linguists will tell you is that repetition matters. You know, if somebody hears ‘she wanted to have sex’ five times, that's gonna have more of an impact than ‘she wanted to have sex’ once\ldots In the end, I think this is a process of what are you prioritizing?}''\looseness=-1
\end{quote}

\noindent Other decision points identified during this phase were determining the modality of annotation (e.g., multiple choice vs. free response) (P4); determining whether and how to collect measures of annotator certainty (P1, P5, P6); navigating the tradeoff between writing annotation guidelines that were comprehensive without being overly long (P9)~\cite[see][]{dimara_narratives_2017, gadiraju_clarity_2017, pradhan_search_2022}; designing an optimal interface for collecting annotations, especially when the data to be annotated was multimodal (P7--P9); and determining the optimal level of coordination among annotators (e.g., should annotators come to a consensus or independently annotate instances?) (P1, P4, P7). Multiple participants noted that all of these decision points are complicated by the rise of AI annotators, as the extent to which best practices for human annotation can be applied to AI is not yet clear (P1, P2, P5, P7, P10)~\cite[see][]{he_if_2024}.

\paragraph{Annotator Management \& Annotation Collection.} By far the biggest decision point in the annotator management phase, reported by all participants, was deciding how to select the annotator population: should annotators be human experts, human crowdworkers, AI, or some combination thereof? Participants reported having to determine whether annotation required expertise or could be done by crowdworkers (P3), and whether annotation required human perspectives or could be done by AI (P6, P8, P10). Participants also grappled with the fact that, although AI annotators tended to make fewer errors than humans, the errors they did make were often less explainable (P2, P5). They also struggled with resource constraints, such as cost and time, that prevented them from regularly using expert human annotators (P1--P3, P6, P7, P9, P10), who were seen as a gold standard.\looseness=-1

There is a large body of research on managing and incentivizing crowdworkers which deals with questions of how to select an annotator population, how to decide whether annotators should be screened (e.g., based on their prior tasks completed on a crowdwork platform) or tested (e.g., by completing an assessment prior to the start of the task), how to communicate with annotators, and how to determine optimal payment structures to incentivize high-quality annotation while still following fair labor practices~\cite{alabduljabbar_task_2016, alonso_practical_2015, calderon-etal-2025-alternative, callison-burch_creating_2010, daniel_quality_2019, eickhoff_cognitive_2018, hansen_quality_2013, ho_incentivizing_2015, kovashka_crowdsourcing_2016, lease_quality_2011, li_crowdsourced_2017, li_dropping_2019, sabou_corpus_2014, snow_cheap_2008, vaughan_making_2018}. Despite this, participants found themselves struggling with many of these questions, as well as with newer problems like preventing unwanted LLM use by crowdworkers (P4, P9).\looseness=-1

\paragraph{Quality Assessment.} Interview participants generally assessed the quality of annotations in three ways: manually reviewing a subset of annotations to see whether they passed the ``sniff test'' (P1--P5, P8--P10), calculating inter-rater reliability (P1--P3, P6, P7, P9, P10), and calculating agreement between annotations and gold labels (P1, P2, P7, P10). Deciding which approach to use (agreement, gold labels, or other) is a universal challenge. Although most annotation managers calculated and tried to reach high IRR, they also acknowledged that high agreement does not necessarily imply high correctness (P1, P7, P8, P10). Nevertheless, they felt that reaching high IRR was often a necessary task to have their annotations seen as valid by third parties. However, as has been extensively discussed in prior work, it is usually not clear which of many possible IRR metrics to rely on (all have been criticized for resting on unrealistic assumptions), nor is it clear what a threshold for ``good'' IRR even is~\cite{banerjee_beyond_1999, bayerl_what_2011, braun_i_2024, checco_lets_2017, mcdonald_reliability_2019, reidsma_reliability_2008, zhao_assumptions_2013}.\footnote{\citet{zhao_assumptions_2013} identify 22 published IRR metrics (many of which are mathematically equivalent). Popular metrics include percent agreement, Cohen's $\kappa$~\cite{cohen_kappa}, and Krippendorff's $\alpha$~\cite{krippendorff_alpha}.} Consistent with prior work, participants often prioritized agreement even when they felt that it was not the best indicator of annotation quality, a phenomenon that was summarized by P10 as ``\textit{a gap between what is publishable and what a peer reviewer will say [is] good enough, and what we actually all should be doing or considering as a research community.}''\looseness=-1

\paragraph{Quality Improvement \& Adjudication.}
In the last stage of the annotation process, teams must decide whether their annotations are sufficiently high-quality. If not, they must decide how to improve annotations: either by collecting new annotations or by removing lower-quality ones. If teams decide to remove lower-quality annotations, several key questions arise about how to do that filtering: what metric and threshold should be used to indicate low quality? Should filtering be done at the data instance level (so that every annotation for a given data instance is removed) or at the annotator level (so that every annotation produced by a particular annotator is removed)? If teams decide that their data is sufficiently high-quality, they must conduct adjudication, which is the process by which a final annotation (or set of annotations) for a data instance is determined. To perform adjudication, annotation teams must first decide whether to retain all labels for an instance or aggregate them into a single label. During this stage, annotation teams may also consider whether and how to express certainty in their final published dataset. Participants reported that their biggest challenge in this stage was deciding whether or not to aggregate labels: multiple participants reported believing that data instances in their setting were genuinely ambiguous, but nevertheless needing a single gold label for their downstream tasks (P2, P5, P9).\looseness=-1

Prior work suggests many approaches for improving the quality of data annotations, including identifying ``hard'' data instances~\cite{beigman_klebanov_annotator_2009}; identifying ``low-quality'' annotators~\cite{dawid_maximum_1979, gordon_disagreement_2021, hovy_learning_2013}; and returning to earlier stages of the annotation process to improve aspects of task design or annotator management~\cite{alonso_practical_2015, hube_understanding_2019, pradhan_search_2022, rottger_two_2022, sabou_corpus_2014, vaughan_making_2018}. Interestingly, some prior work has also proposed that annotation errors---and disagreement generally---can be leveraged in ML processes for learning models that are robust to noise, and that may outperform models trained on `cleaner' data~\cite{Ali_Zhao_Koenecke_Papakyriakopoulos_2026, basile_we_2021, Peterson_2019_ICCV, plank_problem_2022, reidsma_exploiting_2008, schmarje_is_2022, sheng_get_2008, thebault-spieker_diverse_2023, uma_learning_2021}.\looseness=-1

\subsection{Distinct Sources of Annotation Issues}\label{subsec:results:sources}
\begin{table*}
\centering
\begin{small}
\begin{tabular}{>{\hspace{0pt}}m{0.1\linewidth}>
{\hspace{0pt}}m{0.4\linewidth}>{\hspace{0pt}}m{0.45\linewidth}} 
\toprule
\textbf{Source} & \textbf{Definition} & \textbf{Potential Solutions} \\ 
\midrule
\textbf{Error} & The \textbf{annotator} did not follow instructions & Retrain annotators; give feedback to annotators; filter annotators \\
\midrule
\textbf{Ambiguity} & There is not enough information in the \textbf{annotation instructions} to fully determine the annotation & Improve annotation instructions; refine annotation schema\\
\midrule
\textbf{Impossibility} & There is not enough information in the \textbf{data} to fully determine the annotation & Re-process data; filter data \\
\midrule
\textbf{Subjectivity} & The annotation depends on \textbf{implicit values, beliefs, opinions, or assumptions} & Improve annotation instructions; refine annotation schema \\
\midrule
\textbf{Identity} & The identity of the \textbf{annotator} is not aligned with the desired annotator population & Filter annotators \\
\bottomrule
\end{tabular}
\end{small}
\caption{Sources of annotation issues. The source of an issue determines potential solutions to the issue.}
\label{tab:mismatches}
\end{table*}
Annotation issues can often be identified by finding misalignments between an annotation and a ``ground truth.'' But the action needed to correct an annotation issue depends on the source of that issue. Based on our literature review and interview study, we propose a taxonomy of five distinct sources of annotation issues (Table~\ref{tab:mismatches}). While many of these sources have been identified and discussed by prior work, here, we offer a set of definitions that more clearly distinguish unique sources of issues from one another.\looseness=-1

\textit{Errors} occur when annotators do not follow given instructions (P1, P2, P4--P9). A large body of prior work is devoted to identifying errors that occur when crowdworkers try to `game' annotation tasks~\cite{alonso_challenges_2015, marshall_who_2023}. In our interviews, participants were more likely to think of errors as good-faith mistakes stemming from fatigue (P1, P2, P6, P7) or over-reliance on heuristics-based annotation strategies (e.g., annotating for sentiment based on keywords as opposed to careful reading of text) (P2, P4, P7). \textit{Ambiguity} occurs when annotation instructions do not provide enough information to determine what an annotation should be (P2, P3, P5--P8, P10). Ambiguity has been identified by prior work as a source of low agreement in annotation~\cite{braun_i_2024, geva_are_2019, schwirten_ambiguous_2024}. Unlike prior work, we distinguish ambiguity from \textit{impossibility}, which occurs when data instances themselves do not provide enough information to determine what an annotation should be (P5, P7). \textit{Subjectivity} occurs when an annotation depends on the personal values, feelings, or beliefs of an annotator (P1--P3, P5--P10). Many researchers have identified subjectivity, which also contributes to low agreement, as a major challenge in data annotation~\cite{arhin_ground-truth_2021, ferracane_did_2021, hettiachchi_how_2023, kapania_hunt_2023, pang_auditing_2023, parmar_dont_2023, sap_annotators_2022, schumann_consensus_2023, wang_just_2025, wen-yi_automate_2024}. Subjectivity affects even tasks that may sound objective, such as identifying the skin tone of an individual captured in an image~\cite{barrett_skin_2023}. Finally, we introduce \textit{identity} as a source of annotation issues when the quality of an annotation depends not on the content of the annotation, but on the identity of the annotator (P4, P9). For example, AI-generated annotations are misaligned with a desired ``ground truth'' when they are created for an evaluation task intended to elicit human preferences.\looseness=-1

The cause of the issue affects the solutions that can or should be applied to address it. It is important to have a clear language for understanding issues and their causes---without it, annotation managers may try to correct issues using inappropriate methods. For example, \citet{zhang_making_2025} find that annotation managers often treat all low-agreement annotations as errors, when in fact, large-scale low agreement is likely to indicate that annotations may be ambiguous or impossible. Nevertheless, the annotation managers studied by \citet{zhang_making_2025} responded to these cases by rejecting (and thus declining to pay) large swaths of annotators. In addition to constituting an objectionable labor practice, this action will not solve the problem if annotation tasks are ambiguous or impossible.\looseness=-1

\section{Improving Annotation Reliability and Validity with Measurement Theory}\label{sec:results:measurement}
Despite the fact that annotation issues are often thought of as misalignment between annotations and ``ground truths,'' annotation teams consistently focus on agreement (between multiple annotators on the same data instance) as their primary quality metric. This is a fundamental mismatch that is often justified by the fact that agreement is generally easy to calculate and ``ground truth'' is inevitably not available for all data instances (if it were, annotation would not be necessary). Measurement theory offers a broader vocabulary for evaluating annotation quality in terms not only of whether annotators agree, but of whether annotations are reliable and valid measures of the concepts they are intended to capture. In this section, we draw on measurement theory to identify relevant aspects of reliability and validity, and use findings from our literature review and interview study to show how they can be incorporated into annotation processes.\looseness=-1

\paragraph{Test-Retest Reliability.} While inter-rater reliability is the dominant approach to measuring annotation quality, it is rare to assess test-retest reliability (also referred to as intra-rater reliability), which asks whether annotations are consistent when produced by the same annotator at different points in time. \citet{abercrombie_consistency_2025} recently found that, of more than 80,000 papers in the ACL Anthology, just 56 of them (0.07\%) report measuring test-retest reliability. However, researchers hypothesize that test-retest reliability can be used to identify sources of disagreement in annotation (e.g., distinguishing disagreement caused by subjectivity from disagreement caused by random error)~\cite{abercrombie_consistency_2025, gordon_disagreement_2021}. In subjective annotation tasks, measuring test-retest reliability may thus be a better way of identifying high-quality annotations than measuring IRR~\cite{arhin_ground-truth_2021}. Test-retest reliability also has its challenges, however: due to its relative unpopularity, there are currently no agreed-upon best practices for how to best calculate it (for example, how long should researchers wait between original and subsequent annotations?)~\cite{abercrombie_consistency_2025}.\looseness=-1

Consistent with prior work, very few interview participants reported measuring test-retest reliability. Some seemed unaware that it existed, like P1, who asked: ``\textit{Like, check over your own work? I don't know that that's ever been done.}'' However, two participants reported that test-retest reliability had unique benefits not captured in other forms of reliability (P7, P10). P7 described it as ``\textit{low-hanging fruit for quality [assessment]\ldots of the annotator}.'' P10 argued that assessing test-retest reliability is particularly important for subjective tasks, noting that ``\textit{as you code data\ldots your affinity with the data and your understanding of the construct will change.}''\looseness=-1

\paragraph{Face Validity.} Face validity assessments, in which researchers check whether annotated data instances ``look right'' or pass the ``sniff test'' are common---nearly every interview participant reported engaging in face validity assessments by conducting manual review of a subset of annotated data instances (P1--P5, P8--P10). Prior work has found that this practice is widespread~\cite{harvey-etal-2025-understanding}. However, these face validity assessments are often informal and thus neither documented nor published. In a rare exception describing a small-scale face validity assessment, \citet{harvey_framework_2025} described identifying issues in a crowdsourced corpus of tweets written in both African American English and `Standard' American English~\cite{groenwold_investigating_2020}, and choosing not to use the dataset as a result. Similarly, \citet{grondin-verdon_qualitative_2024} conducted a face validity assessment of a gesture recognition corpus by re-annotating parts of the corpus and comparing their annotations to crowdsourced ones. For larger-scale face validity assessments, researchers may turn to automated methods (e.g., confident learning~\cite{northcutt_pervasive_2021}) that are intended to identify likely annotation errors that can then be manually examined~\cite{ klie_annotation_2023}.\looseness=-1

\paragraph{Content Validity.} Content validity asks whether annotations ``wholly and fully capture the substantive nature'' of the concept that they seek to measure~\cite{jacobs_measurement_2021}. In the context of annotation, content validity is most notably considered in social science research. Social science research, which often conceives of annotation as a qualitative coding task, discusses the importance of establishing clear definitions and coding rules for concepts of interest, and of grounding these definitions in specific theories~\cite{macqueen_codebook_1998}. In contrast, critical scholars have repeatedly pointed out that crowdsourced annotation work and annotation tasks intended for AI/ML systems often do not include clear systematizations or operationalizations of concepts of interest~\cite{blodgett_stereotyping_2021, scheuerman_how_2020}.\looseness=-1

Interview participants, especially those with backgrounds in social sciences, described attempting to establish content validity by using established theories to develop their codebooks or annotation guidelines (P1, P2, P10). P10 described how they interrogated the content validity of their annotation guidelines:\looseness=-1 
\begin{quote} 
``\textit{We actually had raters with an understanding of the\ldots theories. We gave them the codes} [i.e., labels] \textit{and had them rate them for how clear are they, how concise are they, how aligned with the theory are they, how practical are they? And so we have these raw metrics of code quality.}''\looseness=-1
\end{quote}

\paragraph{Convergent Validity.} Convergent validity asks whether annotations correlate with other, valid, annotations of related concepts. Because content validity is rarely considered, assessing convergent validity in annotated data is nearly impossible: it is extremely hard to combine and compare different datasets. For example, in a review of language corpora annotated with information on negation, \citet{jimenez-zafra_corpora_2020} found that one-third of datasets lacked annotation guidelines entirely. The others had incomplete or inconsistent guidelines, making it impossible to combine and compare for the purposes of a convergent validity assessment. \citet{bostan_analysis_2018} had similar findings in a review of corpora annotated for emotion classification.\looseness=-1 

Similarly, interview participants reported that they struggled to use existing annotated data and annotation guidelines, whether or not they had been involved in the creation of the data or guidelines in the first place (P1--P4, P6, P10). A primary reason for this was that participants felt that the concepts being annotated were often not consistently systematized or operationalized. Like prior researchers, they struggled with how to compare or combine different systematizations or operationalizations of related concepts.\looseness=-1

Assessing convergent validity thus requires establishing content validity of annotation processes as a first step. Well-documented annotation processes can also be helpful here: for example, \citet{pmlr-v97-recht19a} were able to conduct a convergent validity assessment of ImageNet~\cite{imagenet} by replicating the original process used to select and label images and then testing the ability of models trained on the original ImageNet data to label the new images.\looseness=-1

\paragraph{Predictive Validity.} Predictive validity asks whether annotations can predict relevant related concepts. It is inadvertently assessed almost by definition for many data annotation projects conducted in computational fields: the explicit goal of annotation projects is often to build a model that performs a downstream prediction task. Thus, a view expressed by some interview participants is that annotations can be thought of as valid if they can be used to build an accurate predictive model (P1, P3). However, it is often challenging to specify what exactly the prediction task should be. For example, \citet{catanzariti_taming_2025} conducted an interview study of individuals annotating videos of faces for affect; although the annotations were intended to measure fine-grained emotional states at specific points in time, practitioners assessed quality based on whether models built using the annotations could predict broader outcomes (``is this person generally sad?'').\looseness=-1

\paragraph{Discriminant Validity.}
Discriminant validity asks whether annotations are too correlated with concepts not related to those that they seek to capture. While interview participants did not report engaging in discriminant validity assessments, prior work suggests opportunities for incorporating discriminant validity assessments into the annotation process. For example, researchers have trained models on annotated data that they then use to predict information that should not be contained in annotations---if the models perform well, this is a sign that annotations have low discriminant validity. For example, \citet{geva_are_2019} find that, in widely-used natural language understanding (NLU) benchmarks, it is possible to predict the IDs of annotators who produced particular examples. This suggests that rather than simply measuring NLU, the annotations are also inadvertently capturing information about individual annotators.\looseness=-1

\paragraph{} Taken together, we recommend that researchers and practitioners working on annotation projects should:
\begin{itemize}
    \item[(a)] Determine the source of an annotation error before taking action to resolve it (see Table~\ref{tab:mismatches}), e.g., by measuring test-retest reliability in addition to IRR (to distinguish ambiguity from subjectivity) or by adding an option for annotators to indicate that an instance cannot be annotated (to distinguish ambiguity from impossibility); 
    \item[(b)] Clearly systematize and operationalize concepts of interest to establish content validity and facilitate interrogations of convergent and discriminant validity;
    \item[(c)] Document and report the results of face and predictive validity assessments that are already informally conducted (clearly specifying or pre-registering the predictive task for predictive validity assessments); and
    \item[(d)] Build infrastructure to support interrogation (see \S\ref{sec:discussion}).
\end{itemize}
\section{Discussion}\label{sec:discussion}
In this work, we presented the results of a qualitative study that is intended to facilitate the correction of issues in data annotation. Here, we outline several avenues for future work that our findings suggest will be particularly fruitful.\looseness=-1

\paragraph{Learning from Qualitative Coding.} Multiple participants reported either that they struggled to identify best practices for their annotation tasks (P1, P7), or that they frequently observed other annotation projects that did not follow what they considered to be well-documented best practices (P9, P10). This is largely due to the fact that data annotation is a task that exists across multiple disciplines which are often not in conversation with each other. For example, P1, whose background is in machine learning, felt that ``\textit{as a team we're learning best practices that already exist\ldots more from the qualitative research community.}'' Similarly, P7, whose background is in data science, (jokingly) described the feeling of realizing that their annotation challenge could be addressed by adopting best practices from another field: ``\textit{political science, apparently, has been doing [pairwise annotation] for a long time and hiding it from everybody else.}'' When computer science and machine learning researchers think of data annotation, they often focus on bringing in best practices from crowdwork and even from prediction tasks. We suggest that best practices from the social sciences could be a valuable source of ideas for improving the reliability and validity of data annotation.\looseness=-1

Furthermore, critical scholarship from the social sciences is the primary avenue through which questions of \textit{consequential validity} of data annotations (i.e., what are the societal impacts of annotations?) are examined. \citet{catanzariti_taming_2025}, \citet{denton_genealogy_2021}, \citet{goldstein_how_2022}, \citet{scheuerman_how_2020}, and \citet{scheuerman_datasets_2021}  argue that annotations ``construct'' ground truth, making subjective or value-laden judgments appear objective and immutable. Thus, annotations can reify social hierarchies and reinforce power imbalances~\cite{smart_discipline_2024}. Scholars have also noted that power imbalances can be reinforced within the annotation supply chain~\cite{disalvo_when_2024, miceli_between_2020}. As crowdsourcing became an increasingly popular annotation paradigm in the early 2010s, \citet{irani_turkopticon_2013} raised the alarm that ``human computation currently relies on worker invisibility.'' Despite the efforts of researchers like \citet{irani_turkopticon_2013}, who introduced the Turkopticon system to allow crowdworkers to share experiences, organize, and engage in mutual aid, and \citet{miceli_methodological_2025}, who proposed Workers’ Inquiry as a Research Methodology to center crowdworkers in AI research, annotation labor remains largely invisible today~\cite{gray2019ghost}. Annotation tasks are often outsourced to low-wage workers located in lower-income countries who are subject to myriad risks, including psychological harm from labeling toxic data as well as privacy risks that are inherent to production tasks~\cite{hao2025empire, le_ludec_problem_2023, shestakofsky_cleaning_2024, shmueli_beyond_2021, smart_discipline_2024}. Annotation managers often view annotators as adversarial parties who may try to ``game'' the system by being paid to provide low-effort, low-quality data~\cite{rothschild_problems_2024}. As a result, annotation managers employ various surveillance and quality control measures to remove human uncertainty and subjectivity from annotations, often rejecting annotations and withholding pay from annotators who do not meet certain standards~\cite{meisner_labor_2024, rothschild_problems_2024, zhang_making_2025}. To improve consequential validity in data annotation it is crucial to rely on critical scholarship that centers workers by emphasizing how leveraging worker errors, worker disagreement, and communication with workers can help improve both labor conditions and data quality~\cite{lin_bias_2023, schaekermann_resolvable_2018,  simons_i_2020}.\looseness=-1

\paragraph{Accounting for AI.} Current practices are being adjusted to account for the growing role of AI in data annotation pipelines~\cite{desai2026validatingllmssocialscience, tan_large_2024}. Researchers have proposed a variety of approaches for combining human judgment with AI precision and consistency in order to improve annotation processes~\cite{calderon-etal-2025-alternative, he_if_2024, schroeder_just_2025, tavakoli_reliable_2025}. An overarching theme is that AI annotators have the potential to prevent common human errors in annotation that are caused by fatigue or inattention~\cite{wen-yi_automate_2024}. At the same time, however, researchers are raising alarms about the potential for AI annotators to replace human perspectives in annotation and potentially render crowdsourced data unusable~\cite{wang_large_2025, westwood_potential_2025}. Research shows that LLMs can produce dramatically different annotations depending on minor prompt variation~\cite{rottger_political_2024}, raising concerns of ``LLM-hacking''~\cite{baumann_large_2025}. Co-annotation between humans and LLMs can also lead to humans adopting LLM annotations, even if humans on their own would annotate differently~\cite{he_prompting_2025}. Exploring how annotation teams may use AI to reduce annotation issues---without introducing new ones---is thus an important avenue for future work~\cite[e.g.,][]{Kumar2026}.\looseness=-1

\paragraph{Building Infrastructure to Support Interrogation.}
Finally, future research should explore practical avenues for interrogating the reliability and validity of annotations. In particular, almost all interview participants (P1--P5, P8--P10) reported conducting manual review of annotations to see if they passed the ``sniff test''---i.e., face validity assessments. However, these assessments were informal: participants did not document or publish their results. Building infrastructure to formalize face validity assessments (e.g., by creating repositories for data users to share and aggregate the results of their manual reviews or mechanisms for the research community to flag low-quality annotations in public datasets) could thus enable the improvement of annotated data at scale, in line with prior work that has shown that datasets can be improved by public efforts to correct errors~\cite{beyer2020imagenet, denton_genealogy_2021, northcutt_pervasive_2021, peychev2023automated}.\looseness=-1

\section{Conclusion}\label{sec:conclusion}
Through a literature review (N=132) and interview study (N=10), we contribute a framework to support the correction of annotation issues. First, we map key decision points throughout the annotation process that can lead to annotation issues. By making these decision points, which are often taken for granted, visible, we show where annotation processes can be changed in order to improve annotations. Second, we argue that researchers and practitioners should treat annotation as a measurement problem. In so doing, we build on prior work that has called for measurement theory to be applied to sociotechnical research~\cite{jacobs_measurement_2021, wallach_position_2025} and provide a framework to ground prior work that has called for researchers to look beyond agreement when evaluating data annotation~\cite{thomas2026modernizinggroundtruthshifts}. We make concrete suggestions for how annotation processes can move beyond agreement to interrogate and improve the reliability and validity of annotations.\looseness=-1

\section*{Acknowledgments}
We thank our anonymous interview participants for their time and candor. We thank Nikhil Garg; Karen Levy; Angelina Wang; and the members of the Future of Learning Lab, the KLEAR Lab, the AI in Policy and Practice group, and the Digital Life Initiative for their thoughtful and constructive feedback. This material is based upon work supported by grants from Apple, Inc. and by the National Science Foundation under Grant Number 2416996. Any opinions, findings, and conclusions or recommendations expressed in this material are those of the authors and should not be interpreted as reflecting the views, policies or position, either expressed or implied, of Apple Inc. or the National Science Foundation. 

\bibliography{references}

@article{messick_validity_1995,
	title = {Validity of psychological assessment: {Validation} of inferences from persons' responses and performances as scientific inquiry into score meaning.},
	volume = {50},
	issn = {1935-990X, 0003-066X},
	shorttitle = {Validity of psychological assessment},
	url = {https://doi.apa.org/doi/10.1037/0003-066X.50.9.741},
	doi = {10.1037/0003-066X.50.9.741},
	language = {en},
	number = {9},
	urldate = {2026-06-16},
	journal = {American Psychologist},
	author = {Messick, Samuel},
	month = sep,
	year = {1995},
	pages = {741--749},
}

@misc{desai2026validatingllmssocialscience,
      title={Validating LLMs in social science: Epistemic threats and emerging norms}, 
      author={Meera Desai and Dallas Card and Abigail Z. Jacobs},
      year={2026},
      eprint={2607.07915},
      archivePrefix={arXiv},
      primaryClass={cs.CY},
      url={https://arxiv.org/abs/2607.07915}, 
}

@article{cronbach_construct_1955,
	title = {Construct validity in psychological tests},
	volume = {52},
	issn = {1939-1455, 0033-2909},
	url = {https://doi.apa.org/doi/10.1037/h0040957},
	doi = {10.1037/h0040957},
	language = {en},
	number = {4},
	urldate = {2026-06-16},
	journal = {Psychological Bulletin},
	author = {Cronbach, Lee J. and Meehl, Paul E.},
	month = jul,
	year = {1955},
	pages = {281--302},
}

@article{cook_current_2006,
	title = {Current {Concepts} in {Validity} and {Reliability} for {Psychometric} {Instruments}: {Theory} and {Application}},
	volume = {119},
	issn = {00029343},
	shorttitle = {Current {Concepts} in {Validity} and {Reliability} for {Psychometric} {Instruments}},
	url = {https://linkinghub.elsevier.com/retrieve/pii/S0002934305010375},
	doi = {10.1016/j.amjmed.2005.10.036},
	language = {en},
	number = {2},
	urldate = {2026-06-29},
	journal = {The American Journal of Medicine},
	author = {Cook, David A. and Beckman, Thomas J.},
	month = feb,
	year = {2006},
	pages = {166.e7--166.e16},
}

@article{SNYDER2019333,
title = {Literature review as a research methodology: An overview and guidelines},
journal = {Journal of Business Research},
volume = {104},
pages = {333-339},
year = {2019},
issn = {0148-2963},
doi = {https://doi.org/10.1016/j.jbusres.2019.07.039},
url = {https://www.sciencedirect.com/science/article/pii/S0148296319304564},
author = {Hannah Snyder},
}

@article{Ali_Zhao_Koenecke_Papakyriakopoulos_2026, title={Operationalizing Pluralistic Values in Large Language Model Alignment Reveals Trade-offs in Safety, Inclusivity, and Model Behavior}, volume={40}, url={https://ojs.aaai.org/index.php/AAAI/article/view/41053}, DOI={10.1609/aaai.v40i44.41053}, abstractNote={Although large language models (LLMs) are increasingly trained using human feedback for safety and alignment with human values, alignment decisions often overlook human social diversity. This study examines how incorporating pluralistic values affects LLM behavior by systematically evaluating demographic variation and design parameters in the alignment pipeline. We collect alignment data from US and German participants (N = 1,095 participants, 27,375 ratings) who rated LLM responses across five dimensions: Toxicity, Emotional Awareness (EA), Sensitivity, Stereotypical Bias, and Helpfulness. We fine-tuned multiple Large Language Models and Large Reasoning Models using preferences from different social groups while varying rating scales, disagreement handling methods, and optimization techniques. The results revealed systematic demographic effects: male participants rated responses 18% less toxic than female participants; conservative and Black participants rated responses 27.9% and 44% higher on EA than liberal and White participants, respectively. Models fine-tuned on group-specific preferences exhibited distinct behaviors. Technical design choices showed strong effects: the preservation of rater disagreement achieved roughly 53% greater toxicity reduction than majority voting, and 5-point scales yielded about 22% more reduction than binary formats; and Direct Preference Optimization (DPO) consistently outperformed Group Relative Policy Optimization (GRPO) in multi-value optimization. These findings represent a preliminary step in answering a critical question: How should alignment balance expert-driven and user-driven signals to ensure both safety and fair representation?}, number={44}, journal={Proceedings of the AAAI Conference on Artificial Intelligence}, author={Ali, Dalia and Zhao, Dora and Koenecke, Allison and Papakyriakopoulos, Orestis}, year={2026}, month={Mar.}, pages={37222–37231} }

@Article{Sloane2026,
author={Sloane, Mona
and Schellmann, Hilke
and Mei, Katelyn Xiaoying
and Choi, Anna Seo Gyeong
and Koenecke, Allison},
title={The case for stakeholder-driven AI auditing in automatic speech recognition},
journal={Nature Machine Intelligence},
year={2026},
month={Apr},
day={01},
volume={8},
number={4},
pages={493-494},
issn={2522-5839},
doi={10.1038/s42256-026-01207-x},
url={https://doi.org/10.1038/s42256-026-01207-x}
}

@Article{Kumar2026,
author={Kumar, Aakriti
and Poungpeth, Nalin
and Yang, Diyi
and Farrell, Erina
and Lambert, Bruce L.
and Groh, Matthew},
title={When large language models are reliable for judging empathic communication},
journal={Nature Machine Intelligence},
year={2026},
month={Feb},
day={01},
volume={8},
number={2},
pages={173-185},
issn={2522-5839},
doi={10.1038/s42256-025-01169-6},
url={https://doi.org/10.1038/s42256-025-01169-6}
}

@book{gray2019ghost,
  title={Ghost Work: How to Stop Silicon Valley from Building a New Global Underclass},
  author={Gray, M.L. and Suri, S.},
  isbn={9781328566249},
  lccn={2018042557},
  url={https://books.google.com/books?id=8AmXDwAAQBAJ},
  year={2019},
  publisher={Houghton Mifflin Harcourt}
}

@inproceedings{
peychev2023automated,
title={Automated Classification of Model Errors on ImageNet},
author={Momchil Peychev and Mark Niklas Mueller and Marc Fischer and Martin Vechev},
booktitle={Thirty-seventh Conference on Neural Information Processing Systems},
year={2023},
url={https://openreview.net/forum?id=zEoP4vzFKy}
}

@misc{beyer2020imagenet,
      title={Are we done with ImageNet?}, 
      author={Lucas Beyer and Olivier J. Hénaff and Alexander Kolesnikov and Xiaohua Zhai and Aäron van den Oord},
      year={2020},
      eprint={2006.07159},
      archivePrefix={arXiv},
      primaryClass={cs.CV},
      url={https://arxiv.org/abs/2006.07159}, 
}

@article{messick_validity_1987,
author = {Messick, Samuel},
title = {VALIDITY},
journal = {ETS Research Report Series},
volume = {1987},
number = {2},
pages = {i-208},
doi = {https://doi.org/10.1002/j.2330-8516.1987.tb00244.x},
url = {https://onlinelibrary.wiley.com/doi/abs/10.1002/j.2330-8516.1987.tb00244.x},
eprint = {https://onlinelibrary.wiley.com/doi/pdf/10.1002/j.2330-8516.1987.tb00244.x},
year = {1987}
}

@book{hao2025empire,
  title={Empire of AI: Dreams and Nightmares in Sam Altman's OpenAI},
  author={Hao, K.},
  isbn={9780593657508},
  lccn={2025935502},
  url={https://books.google.com/books?id=Oqo4EQAAQBAJ},
  year={2025},
  publisher={Penguin Publishing Group}
}

@article{cohen_kappa,
author = {Jacob Cohen},
title ={A Coefficient of Agreement for Nominal Scales},
journal = {Educational and Psychological Measurement},
volume = {20},
number = {1},
pages = {37-46},
year = {1960},
doi = {10.1177/001316446002000104},
URL = {https://doi.org/10.1177/001316446002000104}}

@article{krippendorff_alpha,
author = {Andrew F. Hayes and Klaus Krippendorff},
title = {Answering the Call for a Standard Reliability Measure for Coding Data},
journal = {Communication Methods and Measures},
volume = {1},
number = {1},
pages = {77--89},
year = {2007},
publisher = {Routledge},
doi = {10.1080/19312450709336664},
URL = {https://doi.org/10.1080/19312450709336664}
}

@INPROCEEDINGS{imagenet,
  author={Deng, Jia and Dong, Wei and Socher, Richard and Li, Li-Jia and Kai Li and Li Fei-Fei},
  booktitle={2009 IEEE Conference on Computer Vision and Pattern Recognition}, 
  title={ImageNet: A large-scale hierarchical image database}, 
  year={2009},
  volume={},
  number={},
  pages={248-255},
  doi={10.1109/CVPR.2009.5206848}}

@inproceedings{calderon-etal-2025-alternative,
    title = "The Alternative Annotator Test for {LLM}-as-a-Judge: How to Statistically Justify Replacing Human Annotators with {LLM}s",
    author = "Calderon, Nitay  and
      Reichart, Roi  and
      Dror, Rotem",
    editor = "Che, Wanxiang  and
      Nabende, Joyce  and
      Shutova, Ekaterina  and
      Pilehvar, Mohammad Taher",
    booktitle = "Proceedings of the 63rd Annual Meeting of the Association for Computational Linguistics (Volume 1: Long Papers)",
    month = jul,
    year = "2025",
    address = "Vienna, Austria",
    publisher = "Association for Computational Linguistics",
    url = "https://aclanthology.org/2025.acl-long.782/",
    doi = "10.18653/v1/2025.acl-long.782",
    pages = "16051--16081",
    ISBN = "979-8-89176-251-0",
}

@article{westwood_potential_2025,
author = {Sean J. Westwood },
title = {The potential existential threat of large language models to online survey research},
journal = {Proceedings of the National Academy of Sciences},
volume = {122},
number = {47},
pages = {e2518075122},
year = {2025},
doi = {10.1073/pnas.2518075122},
URL = {https://www.pnas.org/doi/abs/10.1073/pnas.2518075122},
eprint = {https://www.pnas.org/doi/pdf/10.1073/pnas.2518075122}}

@InProceedings{pmlr-v97-recht19a,
  title = 	 {Do {I}mage{N}et Classifiers Generalize to {I}mage{N}et?},
  author =       {Recht, Benjamin and Roelofs, Rebecca and Schmidt, Ludwig and Shankar, Vaishaal},
  booktitle = 	 {Proceedings of the 36th International Conference on Machine Learning},
  pages = 	 {5389--5400},
  year = 	 {2019},
  editor = 	 {Chaudhuri, Kamalika and Salakhutdinov, Ruslan},
  volume = 	 {97},
  series = 	 {Proceedings of Machine Learning Research},
  month = 	 {09--15 Jun},
  publisher =    {PMLR},
  url = 	 {https://proceedings.mlr.press/v97/recht19a.html},
}

@InProceedings{Peterson_2019_ICCV,
author = {Peterson, Joshua C. and Battleday, Ruairidh M. and Griffiths, Thomas L. and Russakovsky, Olga},
title = {Human Uncertainty Makes Classification More Robust},
booktitle = {Proceedings of the IEEE/CVF International Conference on Computer Vision (ICCV)},
month = {October},
year = {2019}
}

@inbook{gosciak_scrutinizing_2026,
author = {Gosciak, Jennah and Boyce, Luke and Wang, Angelina and Koenecke, Allison},
title = {Scrutinizing Index-Based Risk Assessments: A Case Study in NYC Decision-making for Heat Emergency Management},
year = {2026},
isbn = {9798400725968},
publisher = {Association for Computing Machinery},
address = {New York, NY, USA},
url = {https://doi.org/10.1145/3805689.3812405},
booktitle = {Proceedings of the 2026 ACM Conference on Fairness, Accountability, and Transparency},
pages = {6659–6701},
numpages = {43}
}

@inproceedings{caine,
author = {Caine, Kelly},
title = {Local Standards for Sample Size at CHI},
year = {2016},
isbn = {9781450333627},
publisher = {Association for Computing Machinery},
address = {New York, NY, USA},
url = {https://doi.org/10.1145/2858036.2858498},
doi = {10.1145/2858036.2858498},
booktitle = {Proceedings of the 2016 CHI Conference on Human Factors in Computing Systems},
pages = {981–992},
numpages = {12},
location = {San Jose, California, USA},
series = {CHI '16}
}

@inproceedings{harvey-etal-2025-understanding,
    title = "Understanding and Meeting Practitioner Needs When Measuring Representational Harms Caused by {LLM}-Based Systems",
    author = "Harvey, Emma  and
      Sheng, Emily  and
      Blodgett, Su Lin  and
      Chouldechova, Alexandra  and
      Garcia-Gathright, Jean  and
      Olteanu, Alexandra  and
      Wallach, Hanna",
    editor = "Che, Wanxiang  and
      Nabende, Joyce  and
      Shutova, Ekaterina  and
      Pilehvar, Mohammad Taher",
    booktitle = "Findings of the Association for Computational Linguistics: ACL 2025",
    month = jul,
    year = "2025",
    address = "Vienna, Austria",
    publisher = "Association for Computational Linguistics",
    url = "https://aclanthology.org/2025.findings-acl.947/",
    doi = "10.18653/v1/2025.findings-acl.947",
    pages = "18423--18440",
    ISBN = "979-8-89176-256-5"
}

@inproceedings{hays_simplistic_2023,
	address = {Austin TX USA},
	title = {Simplistic {Collection} and {Labeling} {Practices} {Limit} the {Utility} of {Benchmark} {Datasets} for {Twitter} {Bot} {Detection}},
	isbn = {978-1-4503-9416-1},
	url = {https://dl.acm.org/doi/10.1145/3543507.3583214},
	doi = {10.1145/3543507.3583214},
	language = {en},
	urldate = {2025-09-22},
	booktitle = {Proceedings of the {ACM} {Web} {Conference} 2023},
	publisher = {ACM},
	author = {Hays, Chris and Schutzman, Zachary and Raghavan, Manish and Walk, Erin and Zimmer, Philipp},
	month = apr,
	year = {2023},
	pages = {3660--3669},
}

@article{hettiachchi_how_2023,
	title = {How {Crowd} {Worker} {Factors} {Influence} {Subjective} {Annotations}: {A} {Study} of {Tagging} {Misogynistic} {Hate} {Speech} in {Tweets}},
	volume = {11},
	issn = {2769-1349, 2769-1330},
	shorttitle = {How {Crowd} {Worker} {Factors} {Influence} {Subjective} {Annotations}},
	url = {https://ojs.aaai.org/index.php/HCOMP/article/view/27546},
	doi = {10.1609/hcomp.v11i1.27546},
	number = {1},
	urldate = {2025-09-22},
	journal = {Proceedings of the AAAI Conference on Human Computation and Crowdsourcing},
	author = {Hettiachchi, Danula and Holcombe-James, Indigo and Livingstone, Stephanie and De Silva, Anjalee and Lease, Matthew and Salim, Flora D. and Sanderson, Mark},
	month = nov,
	year = {2023},
	pages = {38--50},
}

@inproceedings{thebault-spieker_diverse_2023,
	address = {Chicago IL USA},
	title = {Diverse {Perspectives} {Can} {Mitigate} {Political} {Bias} in {Crowdsourced} {Content} {Moderation}},
	isbn = {979-8-4007-0192-4},
	url = {https://dl.acm.org/doi/10.1145/3593013.3594080},
	doi = {10.1145/3593013.3594080},
	language = {en},
	urldate = {2025-10-21},
	booktitle = {2023 {ACM} {Conference} on {Fairness} {Accountability} and {Transparency}},
	publisher = {ACM},
	author = {Thebault-Spieker, Jacob and Venkatagiri, Sukrit and Mine, Naomi and Luther, Kurt},
	month = jun,
	year = {2023},
	pages = {1280--1291},
}

@inproceedings{tavakoli_reliable_2025,
	address = {Padua Italy},
	title = {Reliable {Annotations} with {Less} {Effort}: {Evaluating} {LLM}-{Human} {Collaboration} in {Search} {Clarifications}},
	isbn = {979-8-4007-1861-8},
	shorttitle = {Reliable {Annotations} with {Less} {Effort}},
	url = {https://dl.acm.org/doi/10.1145/3731120.3744574},
	doi = {10.1145/3731120.3744574},
	language = {en},
	urldate = {2025-10-21},
	booktitle = {Proceedings of the 2025 {International} {ACM} {SIGIR} {Conference} on {Innovative} {Concepts} and {Theories} in {Information} {Retrieval} ({ICTIR})},
	publisher = {ACM},
	author = {Tavakoli, Leila and Zamani, Hamed},
	month = jul,
	year = {2025},
	pages = {92--102},
}

@incollection{goldstein_how_2022,
	address = {Lincoln},
	title = {How {Forest} {Became} {Data}: {The} {Remaking} of {Ground}-{Truth} in {Indonesia}},
	isbn = {978-1-4962-1715-8 978-1-4962-3278-6},
	language = {eng},
	booktitle = {The nature of data: infrastructures, environments, politics},
	publisher = {University of Nebraska Press},
	author = {Lin, Cindy},
	editor = {Goldstein, Jenny and Nost, Eric and {Project Muse}},
	year = {2022},
}

@article{disalvo_when_2024,
	series = {{CSCW}},
	title = {When {Workers} {Want} to {Say} {No}: {A} {View} into {Critical} {Consciousness} and {Workplace} {Democracy} in {Data} {Work}},
	volume = {8},
	issn = {2573-0142},
	shorttitle = {When {Workers} {Want} to {Say} {No}},
	url = {https://dl.acm.org/doi/10.1145/3637433},
	doi = {10.1145/3637433},
	language = {en},
	number = {CSCW1},
	urldate = {2025-09-22},
	journal = {Proceedings of the ACM on Human-Computer Interaction},
	author = {DiSalvo, Carl and Rothschild, Annabel and Schenck, Lara L. and Shapiro, Ben Rydal and DiSalvo, Betsy},
	month = apr,
	year = {2024},
	pages = {1--24},
}

@article{miceli_methodological_2025,
	series = {{AIES}},
	title = {Methodological {Considerations} for {Centering} {Workers}’ {Epistemic} {Authority} in {AI} {Research}},
	volume = {8},
	issn = {3065-8365},
	url = {https://ojs.aaai.org/index.php/AIES/article/view/36667},
	doi = {10.1609/aies.v8i2.36667},
	number = {2},
	urldate = {2025-10-24},
	journal = {Proceedings of the AAAI/ACM Conference on AI, Ethics, and Society},
	author = {Miceli, Milagros and Dinika, Adio-Adet and Kauffman, Krystal and Salim Wagner, Camilla and Sachenbacher, Laurenz and Hanna, Alex and Gebru, Timnit},
	month = oct,
	year = {2025},
	pages = {1698--1710},
}

@article{eriksson_can_2025,
	series = {{AIES}},
	title = {Can {We} {Trust} {AI} {Benchmarks}? {An} {Interdisciplinary} {Review} of {Current} {Issues} in {AI} {Evaluation}},
	volume = {8},
	issn = {3065-8365},
	shorttitle = {Can {We} {Trust} {AI} {Benchmarks}?},
	url = {https://ojs.aaai.org/index.php/AIES/article/view/36595},
	doi = {10.1609/aies.v8i1.36595},
	number = {1},
	urldate = {2025-10-24},
	journal = {Proceedings of the AAAI/ACM Conference on AI, Ethics, and Society},
	author = {Eriksson, Maria and Purificato, Erasmo and Noroozian, Arman and Vinagre, João and Chaslot, Guillaume and Gomez, Emilia and Fernandez-Llorca, David},
	month = oct,
	year = {2025},
	pages = {850--864},
}

@incollection{jackman_measurement_2008,
    author = {Jackman, Simon},
    isbn = {9780199286546},
    title = { Measurement},
    booktitle = {The Oxford Handbook of Political Methodology},
    publisher = {Oxford University Press},
    year = {2008},
    month = {08},
    doi = {10.1093/oxfordhb/9780199286546.003.0006},
    url = {https://doi.org/10.1093/oxfordhb/9780199286546.003.0006}
}

@article{paullada_data_2021,
	series = {Patterns},
	title = {Data and its (dis)contents: {A} survey of dataset development and use in machine learning research},
	volume = {2},
	issn = {2666-3899},
	shorttitle = {Data and its (dis)contents},
	url = {https://www.sciencedirect.com/science/article/pii/S2666389921001847},
	doi = {10.1016/j.patter.2021.100336},
	language = {en},
	number = {11},
	urldate = {2023-04-07},
	journal = {Patterns},
	author = {Paullada, Amandalynne and Raji, Inioluwa Deborah and Bender, Emily M. and Denton, Emily and Hanna, Alex},
	month = nov,
	year = {2021},
	pages = {100336},
}

@article{gebru_datasheets_2021,
	title = {Datasheets for datasets},
	volume = {64},
	issn = {0001-0782, 1557-7317},
	url = {https://dl.acm.org/doi/10.1145/3458723},
	doi = {10.1145/3458723},
	language = {en},
	number = {12},
	urldate = {2023-04-07},
	journal = {Communications of the ACM},
	author = {Gebru, Timnit and Morgenstern, Jamie and Vecchione, Briana and Vaughan, Jennifer Wortman and Wallach, Hanna and Iii, Hal Daumé and Crawford, Kate},
	month = dec,
	year = {2021},
	pages = {86--92},
}

@inproceedings{barrett_skin_2023,
	address = {New York, NY, USA},
	series = {{FAccT}},
	title = {Skin {Deep}: {Investigating} {Subjectivity} in {Skin} {Tone} {Annotations} for {Computer} {Vision} {Benchmark} {Datasets}},
	isbn = {979-8-4007-0192-4},
	shorttitle = {Skin {Deep}},
	url = {https://dl.acm.org/doi/10.1145/3593013.3594114},
	doi = {10.1145/3593013.3594114},
	urldate = {2023-06-13},
	booktitle = {Proceedings of the 2023 {ACM} {Conference} on {Fairness}, {Accountability}, and {Transparency}},
	publisher = {Association for Computing Machinery},
	author = {Barrett, Teanna and Chen, Quanze and Zhang, Amy},
	month = jun,
	year = {2023},
	pages = {1757--1771},
}

@inproceedings{pang_auditing_2023,
	address = {New York, NY, USA},
	series = {{FAccT}},
	title = {Auditing {Cross}-{Cultural} {Consistency} of {Human}-{Annotated} {Labels} for {Recommendation} {Systems}},
	isbn = {979-8-4007-0192-4},
	url = {https://dl.acm.org/doi/10.1145/3593013.3594098},
	doi = {10.1145/3593013.3594098},
	urldate = {2023-06-13},
	booktitle = {Proceedings of the 2023 {ACM} {Conference} on {Fairness}, {Accountability}, and {Transparency}},
	publisher = {Association for Computing Machinery},
	author = {Pang, Rock Yuren and Cenatempo, Jack and Graham, Franklyn and Kuehn, Bridgette and Whisenant, Maddy and Botchway, Portia and Stone Perez, Katie and Koenecke, Allison},
	month = jun,
	year = {2023},
	pages = {1531--1552},
}

@inproceedings{jacobs_measurement_2021,
	address = {New York, NY, USA},
	series = {{FAccT}},
	title = {Measurement and {Fairness}},
	isbn = {978-1-4503-8309-7},
	url = {https://dl.acm.org/doi/10.1145/3442188.3445901},
	doi = {10.1145/3442188.3445901},
	urldate = {2023-04-06},
	booktitle = {Proceedings of the 2021 {ACM} {Conference} on {Fairness}, {Accountability}, and {Transparency}},
	publisher = {Association for Computing Machinery},
	author = {Jacobs, Abigail Z. and Wallach, Hanna},
	month = mar,
	year = {2021},
	pages = {375--385},
}

@inproceedings{fabris_tackling_2022,
	address = {Arlington VA USA},
	series = {{EAAMO}},
	title = {Tackling {Documentation} {Debt}: {A} {Survey} on {Algorithmic} {Fairness} {Datasets}},
	isbn = {978-1-4503-9477-2},
	shorttitle = {Tackling {Documentation} {Debt}},
	url = {https://dl.acm.org/doi/10.1145/3551624.3555286},
	doi = {10.1145/3551624.3555286},
	language = {en},
	urldate = {2024-01-26},
	booktitle = {Equity and {Access} in {Algorithms}, {Mechanisms}, and {Optimization}},
	publisher = {ACM},
	author = {Fabris, Alessandro and Messina, Stefano and Silvello, Gianmaria and Susto, Gian Antonio},
	month = oct,
	year = {2022},
	pages = {1--13},
}

@article{scheuerman_datasets_2021,
	series = {{CSCW}},
	title = {Do {Datasets} {Have} {Politics}? {Disciplinary} {Values} in {Computer} {Vision} {Dataset} {Development}},
	volume = {5},
	issn = {2573-0142},
	shorttitle = {Do {Datasets} {Have} {Politics}?},
	url = {https://dl.acm.org/doi/10.1145/3476058},
	doi = {10.1145/3476058},
	language = {en},
	number = {CSCW2},
	urldate = {2024-05-28},
	journal = {Proceedings of the ACM on Human-Computer Interaction},
	author = {Scheuerman, Morgan Klaus and Hanna, Alex and Denton, Emily},
	month = oct,
	year = {2021},
	pages = {1--37},
}

@inproceedings{blodgett_stereotyping_2021,
	address = {Online},
	series = {{ACL}},
	title = {Stereotyping {Norwegian} {Salmon}: {An} {Inventory} of {Pitfalls} in {Fairness} {Benchmark} {Datasets}},
	shorttitle = {Stereotyping {Norwegian} {Salmon}},
	url = {https://aclanthology.org/2021.acl-long.81},
	doi = {10.18653/v1/2021.acl-long.81},
	language = {en},
	urldate = {2024-05-30},
	booktitle = {Proceedings of the 59th {Annual} {Meeting} of the {Association} for {Computational} {Linguistics} and the 11th {International} {Joint} {Conference} on {Natural} {Language} {Processing} ({Volume} 1: {Long} {Papers})},
	publisher = {Association for Computational Linguistics},
	author = {Blodgett, Su Lin and Lopez, Gilsinia and Olteanu, Alexandra and Sim, Robert and Wallach, Hanna},
	month = aug,
	year = {2021},
	pages = {1004--1015},
}

@article{braun_reflecting_2019,
	title = {Reflecting on reflexive thematic analysis},
	volume = {11},
	issn = {2159-676X, 2159-6778},
	url = {https://www.tandfonline.com/doi/full/10.1080/2159676X.2019.1628806},
	doi = {10.1080/2159676X.2019.1628806},
	language = {en},
	number = {4},
	urldate = {2024-07-22},
	journal = {Qualitative Research in Sport, Exercise and Health},
	author = {Braun, Virginia and Clarke, Victoria},
	month = aug,
	year = {2019},
	pages = {589--597},
}

@article{braun_using_2006,
	title = {Using thematic analysis in psychology},
	volume = {3},
	issn = {1478-0887, 1478-0895},
	url = {http://www.tandfonline.com/doi/abs/10.1191/1478088706qp063oa},
	doi = {10.1191/1478088706qp063oa},
	language = {en},
	number = {2},
	urldate = {2024-07-22},
	journal = {Qualitative Research in Psychology},
	author = {Braun, Virginia and Clarke, Victoria},
	month = jan,
	year = {2006},
	pages = {77--101},
}

@inproceedings{zhao_position_2024,
	address = {Vienna, Austria},
	series = {{ICML}},
	title = {Position: {Measure} {Dataset} {Diversity}, {Don}'t {Just} {Claim} {It}},
	copyright = {Creative Commons Attribution 4.0 International},
	url = {https://arxiv.org/abs/2407.08188},
	doi = {10.48550/ARXIV.2407.08188},
	urldate = {2024-07-15},
	booktitle = {The {Forty}-first {International} {Conference} on {Machine} {Learning}},
	author = {Zhao, Dora and Andrews, Jerone T. A. and Papakyriakopoulos, Orestis and Xiang, Alice},
	month = jul,
	year = {2024},
	note = {Version Number: 1},
}

@inproceedings{sambasivan_everyone_2021,
	address = {Yokohama Japan},
	series = {{CHI}},
	title = {“{Everyone} wants to do the model work, not the data work”: {Data} {Cascades} in {High}-{Stakes} {AI}},
	isbn = {978-1-4503-8096-6},
	shorttitle = {“{Everyone} wants to do the model work, not the data work”},
	url = {https://dl.acm.org/doi/10.1145/3411764.3445518},
	doi = {10.1145/3411764.3445518},
	language = {en},
	urldate = {2024-11-04},
	booktitle = {Proceedings of the 2021 {CHI} {Conference} on {Human} {Factors} in {Computing} {Systems}},
	publisher = {ACM},
	author = {Sambasivan, Nithya and Kapania, Shivani and Highfill, Hannah and Akrong, Diana and Paritosh, Praveen and Aroyo, Lora M},
	month = may,
	year = {2021},
	pages = {1--15},
}

@inproceedings{groenwold_investigating_2020,
	address = {Online},
	series = {{EMNLP}},
	title = {Investigating {African}-{American} {Vernacular} {English} in {Transformer}-{Based} {Text} {Generation}},
	url = {https://www.aclweb.org/anthology/2020.emnlp-main.473},
	doi = {10.18653/v1/2020.emnlp-main.473},
	language = {en},
	urldate = {2024-09-16},
	booktitle = {Proceedings of the 2020 {Conference} on {Empirical} {Methods} in {Natural} {Language} {Processing} ({EMNLP})},
	publisher = {Association for Computational Linguistics},
	author = {Groenwold, Sophie and Ou, Lily and Parekh, Aesha and Honnavalli, Samhita and Levy, Sharon and Mirza, Diba and Wang, William Yang},
	month = nov,
	year = {2020},
	pages = {5877--5883},
}

@inproceedings{rottger_political_2024,
	address = {Bangkok, Thailand},
	series = {{ACL}},
	title = {Political {Compass} or {Spinning} {Arrow}? {Towards} {More} {Meaningful} {Evaluations} for {Values} and {Opinions} in {Large} {Language} {Models}},
	shorttitle = {Political {Compass} or {Spinning} {Arrow}?},
	url = {https://aclanthology.org/2024.acl-long.816},
	doi = {10.18653/v1/2024.acl-long.816},
	language = {en},
	urldate = {2024-12-01},
	booktitle = {Proceedings of the 62nd {Annual} {Meeting} of the {Association} for {Computational} {Linguistics} ({Volume} 1: {Long} {Papers})},
	publisher = {Association for Computational Linguistics},
	author = {Röttger, Paul and Hofmann, Valentin and Pyatkin, Valentina and Hinck, Musashi and Kirk, Hannah and Schuetze, Hinrich and Hovy, Dirk},
	month = aug,
	year = {2024},
	pages = {15295--15311},
}

@article{wang_large_2025,
	series = {Nature},
	title = {Large language models that replace human participants can harmfully misportray and flatten identity groups},
	volume = {7},
	issn = {2522-5839},
	url = {https://www.nature.com/articles/s42256-025-00986-z},
	doi = {10.1038/s42256-025-00986-z},
	language = {en},
	number = {3},
	urldate = {2025-03-28},
	journal = {Nature Machine Intelligence},
	author = {Wang, Angelina and Morgenstern, Jamie and Dickerson, John P.},
	month = feb,
	year = {2025},
	pages = {400--411},
}

@inproceedings{rottger_two_2022,
    title = "Two Contrasting Data Annotation Paradigms for Subjective {NLP} Tasks",
    author = {R{\"o}ttger, Paul  and
      Vidgen, Bertie  and
      Hovy, Dirk  and
      Pierrehumbert, Janet},
    editor = "Carpuat, Marine  and
      de Marneffe, Marie-Catherine  and
      Meza Ruiz, Ivan Vladimir",
    booktitle = "Proceedings of the 2022 Conference of the North American Chapter of the Association for Computational Linguistics: Human Language Technologies",
    month = jul,
    year = "2022",
    address = "Seattle, United States",
    publisher = "Association for Computational Linguistics",
    url = "https://aclanthology.org/2022.naacl-main.13/",
    doi = "10.18653/v1/2022.naacl-main.13",
    pages = "175--190"
}

@article{lin_bias_2023,
	series = {{CSCW}},
	title = {From {Bias} to {Repair}: {Error} as a {Site} of {Collaboration} and {Negotiation} in {Applied} {Data} {Science} {Work}},
	volume = {7},
	issn = {2573-0142},
	shorttitle = {From {Bias} to {Repair}},
	url = {https://dl.acm.org/doi/10.1145/3579607},
	doi = {10.1145/3579607},
	language = {en},
	number = {CSCW1},
	urldate = {2025-04-10},
	journal = {Proceedings of the ACM on Human-Computer Interaction},
	author = {Lin, Cindy Kaiying and Jackson, Steven J.},
	month = apr,
	year = {2023},
	pages = {1--32},
}

@inproceedings{torralba_unbiased_2011,
	address = {Colorado Springs, CO, USA},
	series = {{CVPR}},
	title = {Unbiased look at dataset bias},
	isbn = {978-1-4577-0394-2},
	url = {http://ieeexplore.ieee.org/document/5995347/},
	doi = {10.1109/CVPR.2011.5995347},
	urldate = {2025-04-15},
	booktitle = {{CVPR} 2011},
	publisher = {IEEE},
	author = {Torralba, Antonio and Efros, Alexei A.},
	month = jun,
	year = {2011},
	pages = {1521--1528},
}

@article{scheuerman_how_2020,
	series = {{CSCW}},
	title = {How {We}'ve {Taught} {Algorithms} to {See} {Identity}: {Constructing} {Race} and {Gender} in {Image} {Databases} for {Facial} {Analysis}},
	volume = {4},
	issn = {2573-0142},
	shorttitle = {How {We}'ve {Taught} {Algorithms} to {See} {Identity}},
	url = {https://dl.acm.org/doi/10.1145/3392866},
	doi = {10.1145/3392866},
	language = {en},
	number = {CSCW1},
	urldate = {2025-04-15},
	journal = {Proceedings of the ACM on Human-Computer Interaction},
	author = {Scheuerman, Morgan Klaus and Wade, Kandrea and Lustig, Caitlin and Brubaker, Jed R.},
	month = may,
	year = {2020},
	pages = {1--35},
}

@inproceedings{zhang_making_2025,
	address = {Yokohama Japan},
	series = {{CHI}},
	title = {The {Making} of {Performative} {Accuracy} in {AI} {Training}: {Precision} {Labor} and {Its} {Consequences}},
	isbn = {979-8-4007-1394-1},
	shorttitle = {The {Making} of {Performative} {Accuracy} in {AI} {Training}},
	url = {https://dl.acm.org/doi/10.1145/3706598.3713112},
	doi = {10.1145/3706598.3713112},
	language = {en},
	urldate = {2025-05-05},
	booktitle = {Proceedings of the 2025 {CHI} {Conference} on {Human} {Factors} in {Computing} {Systems}},
	publisher = {ACM},
	author = {Zhang, Ben Zefeng and Yang, Tianling and Miceli, Milagros and Haimson, Oliver L. and Thomas, Michaelanne},
	month = apr,
	year = {2025},
	pages = {1--19},
}

@inproceedings{harvey_framework_2025,
	address = {Athens Greece},
	series = {{FAccT}},
	title = {A {Framework} for {Auditing} {Chatbots} for {Dialect}-{Based} {Quality}-of-{Service} {Harms}},
	isbn = {979-8-4007-1482-5},
	url = {https://dl.acm.org/doi/10.1145/3715275.3732137},
	doi = {10.1145/3715275.3732137},
	language = {en},
	urldate = {2025-07-01},
	booktitle = {Proceedings of the 2025 {ACM} {Conference} on {Fairness}, {Accountability}, and {Transparency}},
	publisher = {ACM},
	author = {Harvey, Emma and Kizilcec, Rene F. and Koenecke, Allison},
	month = jun,
	year = {2025},
	pages = {2025--2039},
}

@inproceedings{schroeder_just_2025,
	address = {Vienna, Austria},
	series = {{ACL}},
	title = {Just {Put} a {Human} in the {Loop}? {Investigating} {LLM}-{Assisted} {Annotation} for {Subjective} {Tasks}},
	shorttitle = {Just {Put} a {Human} in the {Loop}?},
	url = {https://aclanthology.org/2025.findings-acl.1323},
	doi = {10.18653/v1/2025.findings-acl.1323},
	language = {en},
	urldate = {2025-08-25},
	booktitle = {Findings of the {Association} for {Computational} {Linguistics}: {ACL} 2025},
	publisher = {Association for Computational Linguistics},
	author = {Schroeder, Hope and Roy, Deb and Kabbara, Jad},
	year = {2025},
	pages = {25771--25795},
}

@misc{thomas_beyond_2025,
	title = {Beyond {Agreement}: {Rethinking} {Ground} {Truth} in {Educational} {AI} {Annotation}},
	shorttitle = {Beyond {Agreement}},
	url = {http://arxiv.org/abs/2508.00143},
	doi = {10.48550/arXiv.2508.00143},
	language = {en},
	urldate = {2025-08-25},
	publisher = {arXiv},
	author = {Thomas, Danielle R. and Borchers, Conrad and Koedinger, Kenneth R.},
	month = jul,
	year = {2025},
	note = {arXiv:2508.00143 [cs]},
}

@article{miceli_between_2020,
	series = {{CSCW}},
	title = {Between {Subjectivity} and {Imposition}: {Power} {Dynamics} in {Data} {Annotation} for {Computer} {Vision}},
	volume = {4},
	issn = {2573-0142},
	shorttitle = {Between {Subjectivity} and {Imposition}},
	url = {https://dl.acm.org/doi/10.1145/3415186},
	doi = {10.1145/3415186},
	language = {en},
	number = {CSCW2},
	urldate = {2025-08-25},
	journal = {Proceedings of the ACM on Human-Computer Interaction},
	author = {Miceli, Milagros and Schuessler, Martin and Yang, Tianling},
	month = oct,
	year = {2020},
	pages = {1--25},
}

@article{catanzariti_taming_2025,
	title = {Taming {Affect}: {On} the {Construction} of {Objectivity} in {Data} {Annotation} {Practices}},
	volume = {4},
	issn = {2731-4650, 2731-4669},
	shorttitle = {Taming {Affect}},
	url = {https://link.springer.com/10.1007/s44206-025-00198-3},
	doi = {10.1007/s44206-025-00198-3},
	language = {en},
	number = {2},
	urldate = {2025-08-25},
	journal = {Digital Society},
	author = {Catanzariti, Benedetta},
	month = aug,
	year = {2025},
	pages = {40},
}

@inproceedings{reidsma_exploiting_2008,
	address = {Manchester, United Kingdom},
	title = {Exploiting 'subjective' annotations},
	isbn = {978-1-905593-49-1},
	url = {http://portal.acm.org/citation.cfm?doid=1611628.1611631},
	doi = {10.3115/1611628.1611631},
	language = {en},
	urldate = {2025-08-25},
	booktitle = {Proceedings of the {Workshop} on {Human} {Judgements} in {Computational} {Linguistics} - {HumanJudge} '08},
	publisher = {Association for Computational Linguistics},
	author = {Reidsma, Dennis and Op Den Akker, Rieks},
	year = {2008},
	pages = {8--16},
}

@article{macqueen_codebook_1998,
	title = {Codebook {Development} for {Team}-{Based} {Qualitative} {Analysis}},
	volume = {10},
	copyright = {https://journals.sagepub.com/page/policies/text-and-data-mining-license},
	issn = {1087-822X},
	url = {https://journals.sagepub.com/doi/10.1177/1525822X980100020301},
	doi = {10.1177/1525822X980100020301},
	language = {en},
	number = {2},
	urldate = {2025-08-25},
	journal = {CAM Journal},
	author = {MacQueen, Kathleen M. and McLellan, Eleanor and Kay, Kelly and Milstein, Bobby},
	month = may,
	year = {1998},
	pages = {31--36},
}

@article{bender_data_2018,
	series = {{TACL}},
	title = {Data {Statements} for {Natural} {Language} {Processing}: {Toward} {Mitigating} {System} {Bias} and {Enabling} {Better} {Science}},
	volume = {6},
	issn = {2307-387X},
	shorttitle = {Data {Statements} for {Natural} {Language} {Processing}},
	url = {https://direct.mit.edu/tacl/article/43452},
	doi = {10.1162/tacl_a_00041},
	language = {en},
	urldate = {2025-08-26},
	journal = {Transactions of the Association for Computational Linguistics},
	author = {Bender, Emily M. and Friedman, Batya},
	month = dec,
	year = {2018},
	pages = {587--604},
}

@inproceedings{diaz_crowdworksheets_2022,
	address = {Seoul Republic of Korea},
	series = {{FAccT}},
	title = {{CrowdWorkSheets}: {Accounting} for {Individual} and {Collective} {Identities} {Underlying} {Crowdsourced} {Dataset} {Annotation}},
	isbn = {978-1-4503-9352-2},
	shorttitle = {{CrowdWorkSheets}},
	url = {https://dl.acm.org/doi/10.1145/3531146.3534647},
	doi = {10.1145/3531146.3534647},
	language = {en},
	urldate = {2025-08-26},
	booktitle = {2022 {ACM} {Conference} on {Fairness} {Accountability} and {Transparency}},
	publisher = {ACM},
	author = {Díaz, Mark and Kivlichan, Ian and Rosen, Rachel and Baker, Dylan and Amironesei, Razvan and Prabhakaran, Vinodkumar and Denton, Remi},
	month = jun,
	year = {2022},
	pages = {2342--2351},
}

@incollection{hallinan_dataset_2020,
	title = {The {Dataset} {Nutrition} {Label}: {A} {Framework} {To} {Drive} {Higher} {Data} {Quality} {Standards}},
	isbn = {978-1-5099-3274-0 978-1-5099-3276-4 978-1-5099-3275-7 978-1-5099-3277-1},
	url = {http://www.bloomsburycollections.com/book/data-protection-and-privacy-data-protection-and-democracy},
	doi = {10.5040/9781509932771},
	language = {en},
	urldate = {2025-08-26},
	booktitle = {Data {Protection} and {Privacy}: {Data} {Protection} and {Democracy}},
	publisher = {Hart Publishing},
	author = {Holland, Sarah and Hosny, Ahmed and Newman, Sarah and Joseph, Joshua and Chmielinski, Kasia},
	editor = {Hallinan, Dara and Leenes, Ronald and Gutwirth, Serge and De Hert, Paul},
	year = {2020},
}

@inproceedings{hutchinson_towards_2021,
	address = {Virtual Event Canada},
	series = {{FAccT}},
	title = {Towards {Accountability} for {Machine} {Learning} {Datasets}: {Practices} from {Software} {Engineering} and {Infrastructure}},
	isbn = {978-1-4503-8309-7},
	shorttitle = {Towards {Accountability} for {Machine} {Learning} {Datasets}},
	url = {https://dl.acm.org/doi/10.1145/3442188.3445918},
	doi = {10.1145/3442188.3445918},
	language = {en},
	urldate = {2025-08-26},
	booktitle = {Proceedings of the 2021 {ACM} {Conference} on {Fairness}, {Accountability}, and {Transparency}},
	publisher = {ACM},
	author = {Hutchinson, Ben and Smart, Andrew and Hanna, Alex and Denton, Remi and Greer, Christina and Kjartansson, Oddur and Barnes, Parker and Mitchell, Margaret},
	month = mar,
	year = {2021},
	pages = {560--575},
}

@inproceedings{gordon_disagreement_2021,
	address = {Yokohama Japan},
	series = {{CHI}},
	title = {The {Disagreement} {Deconvolution}: {Bringing} {Machine} {Learning} {Performance} {Metrics} {In} {Line} {With} {Reality}},
	isbn = {978-1-4503-8096-6},
	shorttitle = {The {Disagreement} {Deconvolution}},
	url = {https://dl.acm.org/doi/10.1145/3411764.3445423},
	doi = {10.1145/3411764.3445423},
	language = {en},
	urldate = {2025-08-26},
	booktitle = {Proceedings of the 2021 {CHI} {Conference} on {Human} {Factors} in {Computing} {Systems}},
	publisher = {ACM},
	author = {Gordon, Mitchell L. and Zhou, Kaitlyn and Patel, Kayur and Hashimoto, Tatsunori and Bernstein, Michael S.},
	month = may,
	year = {2021},
	pages = {1--14},
}

@article{braun_i_2024,
	title = {I beg to differ: how disagreement is handled in the annotation of legal machine learning data sets},
	volume = {32},
	issn = {0924-8463, 1572-8382},
	shorttitle = {I beg to differ},
	url = {https://link.springer.com/10.1007/s10506-023-09369-4},
	doi = {10.1007/s10506-023-09369-4},
	language = {en},
	number = {3},
	urldate = {2025-08-26},
	journal = {Artificial Intelligence and Law},
	author = {Braun, Daniel},
	month = sep,
	year = {2024},
	pages = {839--862},
}

@article{beigman_klebanov_annotator_2009,
	series = {{CL}},
	title = {From {Annotator} {Agreement} to {Noise} {Models}},
	volume = {35},
	issn = {0891-2017, 1530-9312},
	url = {https://direct.mit.edu/coli/article/35/4/495-503/2029},
	doi = {10.1162/coli.2009.35.4.35402},
	language = {en},
	number = {4},
	urldate = {2025-08-26},
	journal = {Computational Linguistics},
	author = {Beigman Klebanov, Beata and Beigman, Eyal},
	month = dec,
	year = {2009},
	pages = {495--503},
}

@article{reidsma_reliability_2008,
	series = {{CL}},
	title = {Reliability {Measurement} without {Limits}},
	volume = {34},
	issn = {0891-2017, 1530-9312},
	url = {https://direct.mit.edu/coli/article/34/3/319-326/1986},
	doi = {10.1162/coli.2008.34.3.319},
	language = {en},
	number = {3},
	urldate = {2025-08-26},
	journal = {Computational Linguistics},
	author = {Reidsma, Dennis and Carletta, Jean},
	month = sep,
	year = {2008},
	pages = {319--326},
}

@misc{baumann_large_2025,
	title = {Large {Language} {Model} {Hacking}: {Quantifying} the {Hidden} {Risks} of {Using} {LLMs} for {Text} {Annotation}},
	shorttitle = {Large {Language} {Model} {Hacking}},
	url = {http://arxiv.org/abs/2509.08825},
	doi = {10.48550/arXiv.2509.08825},
	language = {en},
	urldate = {2025-09-16},
	publisher = {arXiv},
	author = {Baumann, Joachim and Röttger, Paul and Urman, Aleksandra and Wendsjö, Albert and Plaza-del-Arco, Flor Miriam and Gruber, Johannes B. and Hovy, Dirk},
	month = sep,
	year = {2025},
	note = {arXiv:2509.08825 [cs]},
}

@article{klie_analyzing_2024,
	series = {{CL}},
	title = {Analyzing {Dataset} {Annotation} {Quality} {Management} in the {Wild}},
	volume = {50},
	issn = {0891-2017, 1530-9312},
	url = {https://direct.mit.edu/coli/article/50/3/817/120233/Analyzing-Dataset-Annotation-Quality-Management-in},
	doi = {10.1162/coli_a_00516},
	language = {en},
	number = {3},
	urldate = {2025-09-17},
	journal = {Computational Linguistics},
	author = {Klie, Jan-Christoph and Castilho, Richard Eckart De and Gurevych, Iryna},
	month = sep,
	year = {2024},
	pages = {817--866},
}

@misc{smart_discipline_2024,
	title = {Discipline and {Label}: {A} {WEIRD} {Genealogy} and {Social} {Theory} of {Data} {Annotation}},
	copyright = {Creative Commons Attribution 4.0 International},
	shorttitle = {Discipline and {Label}},
	url = {https://arxiv.org/abs/2402.06811},
	doi = {10.48550/ARXIV.2402.06811},
	urldate = {2025-09-17},
	publisher = {arXiv},
	author = {Smart, Andrew and Wang, Ding and Monk, Ellis and Díaz, Mark and Kasirzadeh, Atoosa and Van Liemt, Erin and Schmer-Galunder, Sonja},
	year = {2024},
	note = {Version Number: 1},
}

@article{bayerl_what_2011,
	series = {{CL}},
	title = {What {Determines} {Inter}-{Coder} {Agreement} in {Manual} {Annotations}? {A} {Meta}-{Analytic} {Investigation}},
	volume = {37},
	issn = {0891-2017, 1530-9312},
	shorttitle = {What {Determines} {Inter}-{Coder} {Agreement} in {Manual} {Annotations}?},
	url = {https://direct.mit.edu/coli/article/37/4/699-725/2129},
	doi = {10.1162/COLI_a_00074},
	language = {en},
	number = {4},
	urldate = {2025-09-17},
	journal = {Computational Linguistics},
	author = {Bayerl, Petra Saskia and Paul, Karsten Ingmar},
	month = dec,
	year = {2011},
	pages = {699--725},
}

@inproceedings{sabou_corpus_2014,
	address = {Reykjavik, Iceland},
	title = {Corpus {Annotation} through {Crowdsourcing}: {Towards} {Best} {Practice} {Guidelines}},
	url = {https://aclanthology.org/L14-1412/},
	language = {en},
	booktitle = {Proceedings of the {Ninth} {International} {Conference} on {Language} {Resources} and {Evaluation} ({LREC}'14)},
	publisher = {European Language Resources Association (ELRA)},
	author = {Sabou, Marta and Bontcheva, Kalina and Derczynski, Leon and Scharl, Arno},
	month = may,
	year = {2014},
}

@inproceedings{callison-burch_creating_2010,
	address = {Los Angeles},
	title = {Creating {Speech} and {Language} {Data} {With} {Amazon}'s {Mechanical} {Turk}},
	url = {https://aclanthology.org/W10-0701/},
	booktitle = {Proceedings of the {NAACL} {HLT} 2010 {Workshop} on {Creating} {Speech} and {Language} {Data} with {Amazon}’s {Mechanical} {Turk}},
	publisher = {Association for Computational Linguistics},
	author = {Callison-Burch, Chris and Dredze, Mark},
	month = jun,
	year = {2010},
}

@article{daniel_quality_2019,
	title = {Quality {Control} in {Crowdsourcing}: {A} {Survey} of {Quality} {Attributes}, {Assessment} {Techniques}, and {Assurance} {Actions}},
	volume = {51},
	issn = {0360-0300, 1557-7341},
	shorttitle = {Quality {Control} in {Crowdsourcing}},
	url = {https://dl.acm.org/doi/10.1145/3148148},
	doi = {10.1145/3148148},
	language = {en},
	number = {1},
	urldate = {2025-09-17},
	journal = {ACM Computing Surveys},
	author = {Daniel, Florian and Kucherbaev, Pavel and Cappiello, Cinzia and Benatallah, Boualem and Allahbakhsh, Mohammad},
	month = jan,
	year = {2019},
	pages = {1--40},
}

@article{banerjee_beyond_1999,
	title = {Beyond kappa: {A} review of interrater agreement measures},
	volume = {27},
	copyright = {http://onlinelibrary.wiley.com/termsAndConditions\#vor},
	issn = {0319-5724, 1708-945X},
	shorttitle = {Beyond kappa},
	url = {https://onlinelibrary.wiley.com/doi/10.2307/3315487},
	doi = {10.2307/3315487},
	language = {en},
	number = {1},
	urldate = {2025-09-17},
	journal = {Canadian Journal of Statistics},
	author = {Banerjee, Mousumi and Capozzoli, Michelle and McSweeney, Laura and Sinha, Debajyoti},
	month = mar,
	year = {1999},
	pages = {3--23},
}

@inproceedings{parmar_dont_2023,
	address = {Dubrovnik, Croatia},
	series = {{EACL}},
	title = {Don’t {Blame} the {Annotator}: {Bias} {Already} {Starts} in the {Annotation} {Instructions}},
	shorttitle = {Don’t {Blame} the {Annotator}},
	url = {https://aclanthology.org/2023.eacl-main.130},
	doi = {10.18653/v1/2023.eacl-main.130},
	language = {en},
	urldate = {2025-09-17},
	booktitle = {Proceedings of the 17th {Conference} of the {European} {Chapter} of the {Association} for {Computational} {Linguistics}},
	publisher = {Association for Computational Linguistics},
	author = {Parmar, Mihir and Mishra, Swaroop and Geva, Mor and Baral, Chitta},
	year = {2023},
	pages = {1779--1789},
}

@inproceedings{geva_are_2019,
	address = {Hong Kong, China},
	series = {{EMNLP}-{IJCNLP}},
	title = {Are {We} {Modeling} the {Task} or the {Annotator}? {An} {Investigation} of {Annotator} {Bias} in {Natural} {Language} {Understanding} {Datasets}},
	shorttitle = {Are {We} {Modeling} the {Task} or the {Annotator}?},
	url = {https://www.aclweb.org/anthology/D19-1107},
	doi = {10.18653/v1/D19-1107},
	language = {en},
	urldate = {2025-09-17},
	booktitle = {Proceedings of the 2019 {Conference} on {Empirical} {Methods} in {Natural} {Language} {Processing} and the 9th {International} {Joint} {Conference} on {Natural} {Language} {Processing} ({EMNLP}-{IJCNLP})},
	publisher = {Association for Computational Linguistics},
	author = {Geva, Mor and Goldberg, Yoav and Berant, Jonathan},
	year = {2019},
	pages = {1161--1166},
}

@article{uma_learning_2021,
	series = {{JAIR}},
	title = {Learning from {Disagreement}: {A} {Survey}},
	volume = {72},
	issn = {1076-9757},
	shorttitle = {Learning from {Disagreement}},
	url = {https://jair.org/index.php/jair/article/view/12752},
	doi = {10.1613/jair.1.12752},
	urldate = {2025-09-17},
	journal = {Journal of Artificial Intelligence Research},
	author = {Uma, Alexandra N. and Fornaciari, Tommaso and Hovy, Dirk and Paun, Silviu and Plank, Barbara and Poesio, Massimo},
	month = dec,
	year = {2021},
	pages = {1385--1470},
}

@article{zhao_assumptions_2013,
	title = {Assumptions behind {Intercoder} {Reliability} {Indices}},
	volume = {36},
	issn = {2380-8985, 2380-8977},
	url = {https://academic.oup.com/anncom/article/36/1/419/7885574},
	doi = {10.1080/23808985.2013.11679142},
	language = {en},
	number = {1},
	urldate = {2025-09-17},
	journal = {Annals of the International Communication Association},
	author = {Zhao, Xinshu and Liu, Jun S. and Deng, Ke},
	month = jan,
	year = {2013},
	pages = {419--480},
}

@article{checco_lets_2017,
	series = {{HCOMP}},
	title = {Let's {Agree} to {Disagree}: {Fixing} {Agreement} {Measures} for {Crowdsourcing}},
	volume = {5},
	issn = {2769-1349, 2769-1330},
	shorttitle = {Let's {Agree} to {Disagree}},
	url = {https://ojs.aaai.org/index.php/HCOMP/article/view/13306},
	doi = {10.1609/hcomp.v5i1.13306},
	urldate = {2025-09-17},
	journal = {Proceedings of the AAAI Conference on Human Computation and Crowdsourcing},
	author = {Checco, Alessandro and Roitero, Kevin and Maddalena, Eddy and Mizzaro, Stefano and Demartini, Gianluca},
	month = sep,
	year = {2017},
	pages = {11--20},
}

@inproceedings{ho_incentivizing_2015,
	address = {Florence Italy},
	series = {{WWW}},
	title = {Incentivizing {High} {Quality} {Crowdwork}},
	isbn = {978-1-4503-3469-3},
	url = {https://dl.acm.org/doi/10.1145/2736277.2741102},
	doi = {10.1145/2736277.2741102},
	language = {en},
	urldate = {2025-09-17},
	booktitle = {Proceedings of the 24th {International} {Conference} on {World} {Wide} {Web}},
	publisher = {International World Wide Web Conferences Steering Committee},
	author = {Ho, Chien-Ju and Slivkins, Aleksandrs and Suri, Siddharth and Vaughan, Jennifer Wortman},
	month = may,
	year = {2015},
	pages = {419--429},
}

@inproceedings{snow_cheap_2008,
	series = {{EMNLP}},
	title = {Cheap and fast---but is it good?: evaluating non-expert annotations for natural language tasks},
	url = {https://dl.acm.org/doi/10.5555/1613715.1613751},
	doi = {10.5555/1613715.1613751},
	language = {en},
	booktitle = {{EMNLP} '08: {Proceedings} of the {Conference} on {Empirical} {Methods} in {Natural} {Language} {Processing}},
	author = {Snow, Rion and O'Connor, Brendan and Jurafsky, Daniel and Ng, Andrew Y},
	month = oct,
	year = {2008},
}

@inproceedings{ferracane_did_2021,
	address = {Online},
	series = {{NAACL}},
	title = {Did they answer? {Subjective} acts and intents in conversational discourse},
	shorttitle = {Did they answer?},
	url = {https://aclanthology.org/2021.naacl-main.129},
	doi = {10.18653/v1/2021.naacl-main.129},
	language = {en},
	urldate = {2025-09-18},
	booktitle = {Proceedings of the 2021 {Conference} of the {North} {American} {Chapter} of the {Association} for {Computational} {Linguistics}: {Human} {Language} {Technologies}},
	publisher = {Association for Computational Linguistics},
	author = {Ferracane, Elisa and Durrett, Greg and Li, Junyi Jessy and Erk, Katrin},
	year = {2021},
	pages = {1626--1644},
}

@article{klie_annotation_2023,
	series = {{CL}},
	title = {Annotation {Error} {Detection}: {Analyzing} the {Past} and {Present} for a {More} {Coherent} {Future}},
	volume = {49},
	issn = {0891-2017, 1530-9312},
	shorttitle = {Annotation {Error} {Detection}},
	url = {https://direct.mit.edu/coli/article/49/1/157/113280/Annotation-Error-Detection-Analyzing-the-Past-and},
	doi = {10.1162/coli_a_00464},
	language = {en},
	number = {1},
	urldate = {2025-09-18},
	journal = {Computational Linguistics},
	author = {Klie, Jan-Christoph and Webber, Bonnie and Gurevych, Iryna},
	month = mar,
	year = {2023},
	pages = {157--198},
}

@article{shestakofsky_cleaning_2024,
	series = {Big {Data} \& {Society}},
	title = {Cleaning up data work: {Negotiating} meaning, morality, and inequality in a tech startup},
	volume = {11},
	issn = {2053-9517, 2053-9517},
	shorttitle = {Cleaning up data work},
	url = {https://journals.sagepub.com/doi/10.1177/20539517241285372},
	doi = {10.1177/20539517241285372},
	language = {en},
	number = {3},
	urldate = {2025-09-22},
	journal = {Big Data \& Society},
	author = {Shestakofsky, Benjamin},
	month = sep,
	year = {2024},
	pages = {20539517241285372},
}

@article{le_ludec_problem_2023,
	series = {Big {Data} \& {Society}},
	title = {The problem with annotation. {Human} labour and outsourcing between {France} and {Madagascar}},
	volume = {10},
	issn = {2053-9517, 2053-9517},
	url = {http://journals.sagepub.com/doi/10.1177/20539517231188723},
	doi = {10.1177/20539517231188723},
	language = {en},
	number = {2},
	urldate = {2025-09-22},
	journal = {Big Data \& Society},
	author = {Le Ludec, Clément and Cornet, Maxime and Casilli, Antonio A},
	month = jul,
	year = {2023},
	pages = {20539517231188723},
}

@inproceedings{rothschild_problems_2024,
	series = {{AIES}},
	title = {The {Problems} with {Proxies}: {Making} {Data} {Work} {Visible} through {Requester} {Practices}},
	volume = {7},
	shorttitle = {The {Problems} with {Proxies}},
	url = {https://ojs.aaai.org/index.php/AIES/article/view/31721},
	doi = {10.1609/aies.v7i1.31721},
	language = {en},
	urldate = {2025-09-22},
	booktitle = {Proceedings of the {AAAI}/{ACM} {Conference} on {AI}, {Ethics}, and {Society}},
	author = {Rothschild, Annabel and Wang, Ding and Jayakumar Vilvanathan, Niveditha and Wilcox, Lauren and DiSalvo, Carl and DiSalvo, Betsy},
	month = oct,
	year = {2024},
	pages = {1255--1268},
}

@inproceedings{he_if_2024,
	address = {Honolulu HI USA},
	series = {{CHI}},
	title = {If in a {Crowdsourced} {Data} {Annotation} {Pipeline}, a {GPT}-4},
	isbn = {979-8-4007-0330-0},
	url = {https://dl.acm.org/doi/10.1145/3613904.3642834},
	doi = {10.1145/3613904.3642834},
	language = {en},
	urldate = {2025-09-22},
	booktitle = {Proceedings of the {CHI} {Conference} on {Human} {Factors} in {Computing} {Systems}},
	publisher = {ACM},
	author = {He, Zeyu and Huang, Chieh-Yang and Ding, Chien-Kuang Cornelia and Rohatgi, Shaurya and Huang, Ting-Hao Kenneth},
	month = may,
	year = {2024},
	pages = {1--25},
}

@inproceedings{wen-yi_automate_2024,
	series = {{AIES}},
	title = {Automate or {Assist}? {The} {Role} of {Computational} {Models} in {Identifying} {Gendered} {Discourse} in {US} {Capital} {Trial} {Transcripts}},
	volume = {7},
	shorttitle = {Automate or {Assist}?},
	url = {https://ojs.aaai.org/index.php/AIES/article/view/31746},
	doi = {10.1609/aies.v7i1.31746},
	language = {en},
	urldate = {2025-09-22},
	booktitle = {Proceedings of the {AAAI}/{ACM} {Conference} on {AI}, {Ethics}, and {Society}},
	author = {Wen-Yi, Andrea W and Adamson, Kathryn and Greenfield, Nathalie and Goldberg, Rachel and Babcock, Sandra and Mimno, David and Koenecke, Allison},
	month = oct,
	year = {2024},
	pages = {1556--1566},
}

@inproceedings{northcutt_pervasive_2021,
	series = {{NeurIPS}},
	title = {Pervasive {Label} {Errors} in {Test} {Sets} {Destabilize} {Machine} {Learning} {Benchmarks}},
	url = {https://datasets-benchmarks-proceedings.neurips.cc/paper_files/paper/2021/file/f2217062e9a397a1dca429e7d70bc6ca-Paper-round1.pdf},
	language = {en},
	booktitle = {Proceedings of the {Neural} {Information} {Processing} {Systems} {Track} on {Datasets} and {Benchmarks}},
	author = {Northcutt, Curtis G and Athalye, Anish and Mueller, Jonas},
	year = {2021},
}

@misc{schwirten_ambiguous_2024,
	title = {Ambiguous {Annotations}: {When} is a {Pedestrian} not a {Pedestrian}?},
	shorttitle = {Ambiguous {Annotations}},
	url = {http://arxiv.org/abs/2405.08794},
	doi = {10.48550/arXiv.2405.08794},
	language = {en},
	urldate = {2025-09-22},
	publisher = {arXiv},
	author = {Schwirten, Luisa and Scholz, Jannes and Kondermann, Daniel and Keuper, Janis},
	month = may,
	year = {2024},
	note = {arXiv:2405.08794 [cs]},
}

@article{pradhan_search_2022,
	title = {In {Search} of {Ambiguity}: {A} {Three}-{Stage} {Workflow} {Design} to {Clarify} {Annotation} {Guidelines} for {Crowd} {Workers}},
	volume = {5},
	issn = {2624-8212},
	shorttitle = {In {Search} of {Ambiguity}},
	url = {https://www.frontiersin.org/articles/10.3389/frai.2022.828187/full},
	doi = {10.3389/frai.2022.828187},
	urldate = {2025-09-22},
	journal = {Frontiers in Artificial Intelligence},
	author = {Pradhan, Vivek Krishna and Schaekermann, Mike and Lease, Matthew},
	month = may,
	year = {2022},
	pages = {828187},
}

@inproceedings{lease_quality_2011,
	title = {On {Quality} {Control} and {Machine} {Learning} in {Crowdsourcing}},
	url = {https://cdn.aaai.org/ocs/ws/ws0726/3906-16689-1-PB.pdf},
	language = {en},
	booktitle = {Human {Computation}: {Papers} from the 2011 {AAAI} {Workshop} ({WS}-11-11)},
	author = {Lease, Matthew},
	year = {2011},
}

@misc{arhin_ground-truth_2021,
	title = {Ground-{Truth}, {Whose} {Truth}? -- {Examining} the {Challenges} with {Annotating} {Toxic} {Text} {Datasets}},
	shorttitle = {Ground-{Truth}, {Whose} {Truth}?},
	url = {http://arxiv.org/abs/2112.03529},
	doi = {10.48550/arXiv.2112.03529},
	language = {en},
	urldate = {2025-09-22},
	publisher = {arXiv},
	author = {Arhin, Kofi and Baldini, Ioana and Wei, Dennis and Ramamurthy, Karthikeyan Natesan and Singh, Moninder},
	month = dec,
	year = {2021},
	note = {arXiv:2112.03529 [cs]},
}

@inproceedings{irani_turkopticon_2013,
	address = {Paris France},
	series = {{CHI}},
	title = {Turkopticon: interrupting worker invisibility in amazon mechanical turk},
	isbn = {978-1-4503-1899-0},
	shorttitle = {Turkopticon},
	url = {https://dl.acm.org/doi/10.1145/2470654.2470742},
	doi = {10.1145/2470654.2470742},
	language = {en},
	urldate = {2025-09-22},
	booktitle = {Proceedings of the {SIGCHI} {Conference} on {Human} {Factors} in {Computing} {Systems}},
	publisher = {ACM},
	author = {Irani, Lilly C. and Silberman, M. Six},
	month = apr,
	year = {2013},
	pages = {611--620},
}

@article{kovashka_crowdsourcing_2016,
	title = {Crowdsourcing in {Computer} {Vision}},
	volume = {10},
	issn = {1572-2740, 1572-2759},
	url = {http://www.nowpublishers.com/article/Details/CGV-071},
	doi = {10.1561/0600000071},
	language = {en},
	number = {3},
	urldate = {2025-09-22},
	journal = {Foundations and Trends® in Computer Graphics and Vision},
	author = {Kovashka, Adriana and Russakovsky, Olga and Fei-Fei, Li and Grauman, Kristen},
	year = {2016},
	pages = {177--243},
}

@inproceedings{sheng_get_2008,
	address = {Las Vegas Nevada USA},
	series = {{KDD}},
	title = {Get another label? improving data quality and data mining using multiple, noisy labelers},
	isbn = {978-1-60558-193-4},
	shorttitle = {Get another label?},
	url = {https://dl.acm.org/doi/10.1145/1401890.1401965},
	doi = {10.1145/1401890.1401965},
	language = {en},
	urldate = {2025-09-22},
	booktitle = {Proceedings of the 14th {ACM} {SIGKDD} international conference on {Knowledge} discovery and data mining},
	publisher = {ACM},
	author = {Sheng, Victor S. and Provost, Foster and Ipeirotis, Panagiotis G.},
	month = aug,
	year = {2008},
	pages = {614--622},
}

@article{vaughan_making_2018,
	series = {{JMLR}},
	title = {Making {Better} {Use} of the {Crowd}: {How} {Crowdsourcing} {Can} {Advance} {Machine} {Learning} {Research}},
	volume = {18},
	url = {http://jmlr.org/papers/volume18/17-234/17-234.pdf},
	language = {en},
	number = {193},
	journal = {Journal of Machine Learning Research},
	author = {Vaughan, Jennifer Wortman},
	year = {2018},
}

@article{denton_genealogy_2021,
	series = {Big {Data} \& {Society}},
	title = {On the genealogy of machine learning datasets: {A} critical history of {ImageNet}},
	volume = {8},
	issn = {2053-9517, 2053-9517},
	shorttitle = {On the genealogy of machine learning datasets},
	url = {https://journals.sagepub.com/doi/10.1177/20539517211035955},
	doi = {10.1177/20539517211035955},
	language = {en},
	number = {2},
	urldate = {2025-09-23},
	journal = {Big Data \& Society},
	author = {Denton, Emily and Hanna, Alex and Amironesei, Razvan and Smart, Andrew and Nicole, Hilary},
	month = jul,
	year = {2021},
	pages = {20539517211035955},
}

@inproceedings{kapania_hunt_2023,
	address = {Hamburg Germany},
	series = {{CHI}},
	title = {A hunt for the {Snark}: {Annotator} {Diversity} in {Data} {Practices}},
	isbn = {978-1-4503-9421-5},
	shorttitle = {A hunt for the {Snark}},
	url = {https://dl.acm.org/doi/10.1145/3544548.3580645},
	doi = {10.1145/3544548.3580645},
	language = {en},
	urldate = {2025-10-09},
	booktitle = {Proceedings of the 2023 {CHI} {Conference} on {Human} {Factors} in {Computing} {Systems}},
	publisher = {ACM},
	author = {Kapania, Shivani and Taylor, Alex S and Wang, Ding},
	month = apr,
	year = {2023},
	pages = {1--15},
}

@inproceedings{hansen_quality_2013,
	address = {San Antonio Texas USA},
	series = {{CSCW}},
	title = {Quality control mechanisms for crowdsourcing: peer review, arbitration, \& expertise at familysearch indexing},
	isbn = {978-1-4503-1331-5},
	shorttitle = {Quality control mechanisms for crowdsourcing},
	url = {https://dl.acm.org/doi/10.1145/2441776.2441848},
	doi = {10.1145/2441776.2441848},
	language = {en},
	urldate = {2025-10-21},
	booktitle = {Proceedings of the 2013 conference on {Computer} supported cooperative work},
	publisher = {ACM},
	author = {Hansen, Derek L. and Schone, Patrick J. and Corey, Douglas and Reid, Matthew and Gehring, Jake},
	month = feb,
	year = {2013},
	pages = {649--660},
}

@article{alonso_challenges_2015,
	title = {Challenges with {Label} {Quality} for {Supervised} {Learning}},
	volume = {6},
	issn = {1936-1955, 1936-1963},
	url = {https://dl.acm.org/doi/10.1145/2724721},
	doi = {10.1145/2724721},
	language = {en},
	number = {1},
	urldate = {2025-10-21},
	journal = {Journal of Data and Information Quality},
	author = {Alonso, Omar},
	month = mar,
	year = {2015},
	pages = {1--3},
}

@inproceedings{alonso_practical_2015,
	address = {Santiago Chile},
	series = {{SIGIR}},
	title = {Practical {Lessons} for {Gathering} {Quality} {Labels} at {Scale}},
	isbn = {978-1-4503-3621-5},
	url = {https://dl.acm.org/doi/10.1145/2766462.2776778},
	doi = {10.1145/2766462.2776778},
	language = {en},
	urldate = {2025-10-21},
	booktitle = {Proceedings of the 38th {International} {ACM} {SIGIR} {Conference} on {Research} and {Development} in {Information} {Retrieval}},
	publisher = {ACM},
	author = {Alonso, Omar},
	month = aug,
	year = {2015},
	pages = {1089--1092},
}

@inproceedings{alabduljabbar_task_2016,
	address = {Odense Denmark},
	title = {A {Task} {Ontology}-based {Model} for {Quality} {Control} in {Crowdsourcing} {Systems}},
	isbn = {978-1-4503-4455-5},
	url = {https://dl.acm.org/doi/10.1145/2987386.2987413},
	doi = {10.1145/2987386.2987413},
	language = {en},
	urldate = {2025-10-21},
	booktitle = {Proceedings of the {International} {Conference} on {Research} in {Adaptive} and {Convergent} {Systems}},
	publisher = {ACM},
	author = {Alabduljabbar, Reham and Al-Dossari, Hmood},
	month = oct,
	year = {2016},
	pages = {22--28},
}

@inproceedings{gadiraju_clarity_2017,
	address = {Prague Czech Republic},
	title = {Clarity is a {Worthwhile} {Quality}: {On} the {Role} of {Task} {Clarity} in {Microtask} {Crowdsourcing}},
	isbn = {978-1-4503-4708-2},
	shorttitle = {Clarity is a {Worthwhile} {Quality}},
	url = {https://dl.acm.org/doi/10.1145/3078714.3078715},
	doi = {10.1145/3078714.3078715},
	language = {en},
	urldate = {2025-10-21},
	booktitle = {Proceedings of the 28th {ACM} {Conference} on {Hypertext} and {Social} {Media}},
	publisher = {ACM},
	author = {Gadiraju, Ujwal and Yang, Jie and Bozzon, Alessandro},
	month = jul,
	year = {2017},
	pages = {5--14},
}

@inproceedings{marshall_who_2023,
	address = {Austin TX USA},
	series = {{WebSci}},
	title = {Who {Broke} {Amazon} {Mechanical} {Turk}?: {An} {Analysis} of {Crowdsourcing} {Data} {Quality} over {Time}},
	isbn = {979-8-4007-0089-7},
	shorttitle = {Who {Broke} {Amazon} {Mechanical} {Turk}?},
	url = {https://dl.acm.org/doi/10.1145/3578503.3583622},
	doi = {10.1145/3578503.3583622},
	language = {en},
	urldate = {2025-10-21},
	booktitle = {Proceedings of the 15th {ACM} {Web} {Science} {Conference} 2023},
	publisher = {ACM},
	author = {Marshall, Catherine C. and Goguladinne, Partha S.R. and Maheshwari, Mudit and Sathe, Apoorva and Shipman, Frank M.},
	month = apr,
	year = {2023},
	pages = {335--345},
}

@inproceedings{eickhoff_cognitive_2018,
	address = {Marina Del Rey CA USA},
	series = {{WSDM}},
	title = {Cognitive {Biases} in {Crowdsourcing}},
	isbn = {978-1-4503-5581-0},
	url = {https://dl.acm.org/doi/10.1145/3159652.3159654},
	doi = {10.1145/3159652.3159654},
	language = {en},
	urldate = {2025-10-21},
	booktitle = {Proceedings of the {Eleventh} {ACM} {International} {Conference} on {Web} {Search} and {Data} {Mining}},
	publisher = {ACM},
	author = {Eickhoff, Carsten},
	month = feb,
	year = {2018},
	pages = {162--170},
}

@article{li_dropping_2019,
	series = {{CSCW}},
	title = {Dropping the {Baton}?: {Understanding} {Errors} and {Bottlenecks} in a {Crowdsourced} {Sensemaking} {Pipeline}},
	volume = {3},
	issn = {2573-0142},
	shorttitle = {Dropping the {Baton}?},
	url = {https://dl.acm.org/doi/10.1145/3359238},
	doi = {10.1145/3359238},
	language = {en},
	number = {CSCW},
	urldate = {2025-10-21},
	journal = {Proceedings of the ACM on Human-Computer Interaction},
	author = {Li, Tianyi and Manns, Chandler J. and North, Chris and Luther, Kurt},
	month = nov,
	year = {2019},
	pages = {1--26},
}

@inproceedings{hube_understanding_2019,
	address = {Glasgow Scotland Uk},
	series = {{CHI}},
	title = {Understanding and {Mitigating} {Worker} {Biases} in the {Crowdsourced} {Collection} of {Subjective} {Judgments}},
	isbn = {978-1-4503-5970-2},
	url = {https://dl.acm.org/doi/10.1145/3290605.3300637},
	doi = {10.1145/3290605.3300637},
	language = {en},
	urldate = {2025-10-21},
	booktitle = {Proceedings of the 2019 {CHI} {Conference} on {Human} {Factors} in {Computing} {Systems}},
	publisher = {ACM},
	author = {Hube, Christoph and Fetahu, Besnik and Gadiraju, Ujwal},
	month = may,
	year = {2019},
	pages = {1--12},
}

@inproceedings{schmarje_is_2022,
	series = {{NeurIPS}},
	title = {Is one annotation enough? - {A} data-centric image classification benchmark for noisy and ambiguous label estimation},
	volume = {35},
	url = {https://proceedings.neurips.cc/paper_files/paper/2022/file/d6c03035b8bc551f474f040fe8607cab-Paper-Datasets_and_Benchmarks.pdf},
	booktitle = {Advances in {Neural} {Information} {Processing} {Systems}},
	publisher = {Curran Associates, Inc.},
	author = {Schmarje, Lars and Grossmann, Vasco and Zelenka, Claudius and Dippel, Sabine and Kiko, Rainer and Oszust, Mariusz and Pastell, Matti and Stracke, Jenny and Valros, Anna and Volkmann, Nina and Koch, Reinhard},
	editor = {Koyejo, S. and Mohamed, S. and Agarwal, A. and Belgrave, D. and Cho, K. and Oh, A.},
	year = {2022},
	pages = {33215--33232},
}

@inproceedings{schumann_consensus_2023,
	series = {{NeurIPS}},
	title = {Consensus and {Subjectivity} of {Skin} {Tone} {Annotation} for {ML} {Fairness}},
	volume = {36},
	url = {https://proceedings.neurips.cc/paper_files/paper/2023/file/60d25b3210c92f5ba2002a8e1f1adf1c-Paper-Datasets_and_Benchmarks.pdf},
	booktitle = {Advances in {Neural} {Information} {Processing} {Systems}},
	publisher = {Curran Associates, Inc.},
	author = {Schumann, Candice and Olanubi, Femi and Wright, Auriel and Monk, Ellis and Heldreth, Courtney and Ricco, Susanna},
	editor = {Oh, A. and Naumann, T. and Globerson, A. and Saenko, K. and Hardt, M. and Levine, S.},
	year = {2023},
	pages = {30319--30348},
}

@article{schaekermann_resolvable_2018,
	series = {{CSCW}},
	title = {Resolvable vs. {Irresolvable} {Disagreement}: {A} {Study} on {Worker} {Deliberation} in {Crowd} {Work}},
	volume = {2},
	issn = {2573-0142},
	shorttitle = {Resolvable vs. {Irresolvable} {Disagreement}},
	url = {https://dl.acm.org/doi/10.1145/3274423},
	doi = {10.1145/3274423},
	language = {en},
	number = {CSCW},
	urldate = {2025-10-21},
	journal = {Proceedings of the ACM on Human-Computer Interaction},
	author = {Schaekermann, Mike and Goh, Joslin and Larson, Kate and Law, Edith},
	month = nov,
	year = {2018},
	pages = {1--19},
}

@inproceedings{grondin-verdon_qualitative_2024,
	address = {San Jose Costa Rica},
	title = {Qualitative study of gesture annotation corpus : {Challenges} and perspectives},
	isbn = {979-8-4007-0463-5},
	shorttitle = {Qualitative study of gesture annotation corpus},
	url = {https://dl.acm.org/doi/10.1145/3686215.3688820},
	doi = {10.1145/3686215.3688820},
	language = {en},
	urldate = {2025-10-21},
	booktitle = {Companion {Proceedings} of the 26th {International} {Conference} on {Multimodal} {Interaction}},
	publisher = {ACM},
	author = {Grondin-Verdon, Mickaëlla and Caillat, Domitille and Ouni, Slim},
	month = nov,
	year = {2024},
	pages = {147--155},
}

@inproceedings{dimara_narratives_2017,
	address = {Denver Colorado USA},
	series = {{CHI}},
	title = {Narratives in {Crowdsourced} {Evaluation} of {Visualizations}: {A} {Double}-{Edged} {Sword}?},
	isbn = {978-1-4503-4655-9},
	shorttitle = {Narratives in {Crowdsourced} {Evaluation} of {Visualizations}},
	url = {https://dl.acm.org/doi/10.1145/3025453.3025870},
	doi = {10.1145/3025453.3025870},
	language = {en},
	urldate = {2025-10-21},
	booktitle = {Proceedings of the 2017 {CHI} {Conference} on {Human} {Factors} in {Computing} {Systems}},
	publisher = {ACM},
	author = {Dimara, Evanthia and Bezerianos, Anastasia and Dragicevic, Pierre},
	month = may,
	year = {2017},
	pages = {5475--5484},
}

@inproceedings{li_crowdsourced_2017,
	address = {Chicago Illinois USA},
	title = {Crowdsourced {Data} {Management}: {Overview} and {Challenges}},
	isbn = {978-1-4503-4197-4},
	shorttitle = {Crowdsourced {Data} {Management}},
	url = {https://dl.acm.org/doi/10.1145/3035918.3054776},
	doi = {10.1145/3035918.3054776},
	language = {en},
	urldate = {2025-10-21},
	booktitle = {Proceedings of the 2017 {ACM} {International} {Conference} on {Management} of {Data}},
	publisher = {ACM},
	author = {Li, Guoliang and Zheng, Yudian and Fan, Ju and Wang, Jiannan and Cheng, Reynold},
	month = may,
	year = {2017},
	pages = {1711--1716},
}

@article{simons_i_2020,
	series = {{CSCW}},
	title = {"{I} {Hope} {This} {Is} {Helpful}": {Understanding} {Crowdworkers}' {Challenges} and {Motivations} for an {Image} {Description} {Task}},
	volume = {4},
	issn = {2573-0142},
	shorttitle = {"{I} {Hope} {This} {Is} {Helpful}"},
	url = {https://dl.acm.org/doi/10.1145/3415176},
	doi = {10.1145/3415176},
	language = {en},
	number = {CSCW2},
	urldate = {2025-10-21},
	journal = {Proceedings of the ACM on Human-Computer Interaction},
	author = {Simons, Rachel N. and Gurari, Danna and Fleischmann, Kenneth R.},
	month = oct,
	year = {2020},
	pages = {1--26},
}

@article{wang_just_2025,
	series = {{AIES}},
	title = {"{Just} a {Strange} {Pic}": {Rethinking} '{Safety}' in {GenAI} {Image} {Safety} {Annotation} {Tasks} from {Diverse} {Annotators}' {Perspectives}},
	volume = {8},
	issn = {3065-8365},
	shorttitle = {"{Just} a {Strange} {Pic}"},
	url = {https://ojs.aaai.org/index.php/AIES/article/view/36742},
	doi = {10.1609/aies.v8i3.36742},
	number = {3},
	urldate = {2025-10-24},
	journal = {Proceedings of the AAAI/ACM Conference on AI, Ethics, and Society},
	author = {Wang, Ding and Díaz, Mark and Rastogi, Charvi and Davani, Aida and Prabhakaran, Vinodkumar and Mishra, Pushkar and Patel, Roma and Parrish, Alicia and Ashwood, Zoe and Paganini, Michela and Teh, Tian Huey and Rieser, Verena and Aroyo, Lora},
	month = oct,
	year = {2025},
	pages = {2612--2624},
}

@inproceedings{tan_large_2024,
	address = {Miami, Florida, USA},
	series = {{EMNLP}},
	title = {Large {Language} {Models} for {Data} {Annotation} and {Synthesis}: {A} {Survey}},
	shorttitle = {Large {Language} {Models} for {Data} {Annotation} and {Synthesis}},
	url = {https://aclanthology.org/2024.emnlp-main.54},
	doi = {10.18653/v1/2024.emnlp-main.54},
	language = {en},
	urldate = {2025-11-17},
	booktitle = {Proceedings of the 2024 {Conference} on {Empirical} {Methods} in {Natural} {Language} {Processing}},
	publisher = {Association for Computational Linguistics},
	author = {Tan, Zhen and Li, Dawei and Wang, Song and Beigi, Alimohammad and Jiang, Bohan and Bhattacharjee, Amrita and Karami, Mansooreh and Li, Jundong and Cheng, Lu and Liu, Huan},
	year = {2024},
	pages = {930--957},
}

@inproceedings{bostan_analysis_2018,
	address = {Santa Fe, NM, USA},
	series = {{ICCL}},
	title = {An {Analysis} of {Annotated} {Corpora} for {Emotion} {Classification} in {Text}},
	language = {en},
	booktitle = {Proceedings of the 27th {International} {Conference} on {Computational} {Linguistics}},
	publisher = {ACL},
	author = {Bostan, Laura Ana Maria and Klinger, Roman},
	month = aug,
	year = {2018},
}

@article{jimenez-zafra_corpora_2020,
	series = {{CL}},
	title = {Corpora {Annotated} with {Negation}: {An} {Overview}},
	volume = {46},
	issn = {0891-2017, 1530-9312},
	shorttitle = {Corpora {Annotated} with {Negation}},
	url = {https://direct.mit.edu/coli/article/46/1/1-52/93383},
	doi = {10.1162/coli_a_00371},
	language = {en},
	number = {1},
	urldate = {2025-11-17},
	journal = {Computational Linguistics},
	author = {Jiménez-Zafra, Salud María and Morante, Roser and Teresa Martín-Valdivia, María and Ureña-López, L. Alfonso},
	month = mar,
	year = {2020},
	pages = {1--52},
}

@inproceedings{sap_annotators_2022,
	address = {Seattle, United States},
	series = {{NAACL}},
	title = {Annotators with {Attitudes}: {How} {Annotator} {Beliefs} {And} {Identities} {Bias} {Toxic} {Language} {Detection}},
	shorttitle = {Annotators with {Attitudes}},
	url = {https://aclanthology.org/2022.naacl-main.431},
	doi = {10.18653/v1/2022.naacl-main.431},
	language = {en},
	urldate = {2025-11-28},
	booktitle = {Proceedings of the 2022 {Conference} of the {North} {American} {Chapter} of the {Association} for {Computational} {Linguistics}: {Human} {Language} {Technologies}},
	publisher = {Association for Computational Linguistics},
	author = {Sap, Maarten and Swayamdipta, Swabha and Vianna, Laura and Zhou, Xuhui and Choi, Yejin and Smith, Noah},
	year = {2022},
	pages = {5884--5906},
}

@inproceedings{shmueli_beyond_2021,
	address = {Online},
	series = {{NAACL}},
	title = {Beyond {Fair} {Pay}: {Ethical} {Implications} of {NLP} {Crowdsourcing}},
	shorttitle = {Beyond {Fair} {Pay}},
	url = {https://aclanthology.org/2021.naacl-main.295},
	doi = {10.18653/v1/2021.naacl-main.295},
	language = {en},
	urldate = {2025-11-28},
	booktitle = {Proceedings of the 2021 {Conference} of the {North} {American} {Chapter} of the {Association} for {Computational} {Linguistics}: {Human} {Language} {Technologies}},
	publisher = {Association for Computational Linguistics},
	author = {Shmueli, Boaz and Fell, Jan and Ray, Soumya and Ku, Lun-Wei},
	year = {2021},
	pages = {3758--3769},
}

@article{meisner_labor_2024,
	title = {The labor of search engine evaluation: {Making} algorithms more human or humans more algorithmic?},
	volume = {26},
	issn = {1461-4448, 1461-7315},
	shorttitle = {The labor of search engine evaluation},
	url = {https://journals.sagepub.com/doi/10.1177/14614448211063860},
	doi = {10.1177/14614448211063860},
	language = {en},
	number = {2},
	urldate = {2025-12-01},
	journal = {New Media \& Society},
	author = {Meisner, Colten and Duffy, Brooke Erin and Ziewitz, Malte},
	month = feb,
	year = {2024},
	pages = {1018--1033},
}

@article{adcock_measurement_2001,
	title = {Measurement {Validity}: {A} {Shared} {Standard} for {Qualitative} and {Quantitative} {Research}},
	volume = {95},
	copyright = {https://www.cambridge.org/core/terms},
	issn = {0003-0554, 1537-5943},
	shorttitle = {Measurement {Validity}},
	url = {https://www.cambridge.org/core/product/identifier/S0003055401003100/type/journal_article},
	doi = {10.1017/S0003055401003100},
	language = {en},
	number = {3},
	urldate = {2024-05-22},
	journal = {American Political Science Review},
	author = {Adcock, Robert and Collier, David},
	month = sep,
	year = {2001},
	pages = {529--546},
}

\appendix
\section{Semi-Structured Interview Guides}\label{app:interviewguide}
\textit{Prior to scheduling the interview, participants were invited to review study details and provide their written informed consent to participant. Once the interview started, but prior to the start of recording, participants were given a brief overview of the study, invited to ask questions, and asked to provide verbal informed consent to participate in the interview and be recorded.}
\newline\newline\noindent \textit{Because pilot interviews were used to refine the interview guide, some participants were asked a slightly different set of questions, or questions in a slightly different order, from those shown below. As is typical of semi-structured interview studies, some participants were also asked additional follow-ups not listed here.}
\newline\newline\noindent \textbf{Background and Definitions:}
\begin{itemize}
    \item For the purposes of this study, we will be defining data annotation as the process of augmenting data so that it can be used in some downstream task. Does this sound correct to you?
    \item Do you use any other words to refer to data annotation?
\end{itemize}

\noindent \textbf{Participant Experience with Annotation:}
\begin{itemize}
    \item Can you start by briefly listing the annotation projects that you have been involved with?
    \begin{itemize}
        \item \textit{Repeat for as many projects as time permits:} Can you describe the data annotation process on [specific project]?
        \begin{itemize}
            \item Can you describe your role on the project?
            \item What was the data medium and data size?
            \item What was the intended goal or use case?
            \item What did the process of writing annotation instructions look like?
            \item Did you use expert human annotators, crowdworkers, LLMs, or another population of annotators? 
            \item How many annotators did you use? 
            \item How long did the annotation process take? 
            \item What tools or platforms did you use to collect annotations?
            \item To your knowledge, was there a process or tool used to validate data annotations? If so, what was it? 
            \begin{itemize}
                \item At what frequency do you conduct validation? 
                \item Are there specific validation metrics that you use? 
            \end{itemize}
        \end{itemize}
    \end{itemize}
\end{itemize}

\noindent \textbf{Annotation Challenges:}
\begin{itemize}
    \item While working on the data annotation process, what challenges did you face?
    \begin{itemize}
        \item Are there any additional challenges that you observed others involved in the process facing?
    \end{itemize}
    \item \textit{For each challenge:}
    \begin{itemize}
        \item What do you think the cause of [challenge] was? 
        \item What did you do to address [challenge]? 
        \item How, if at all, did this challenge, or the actions you took to address the challenge, affect your data annotation process and the quality of your annotations?
    \end{itemize}
    \item One thing we are particularly interested in is the idea of ``mismatches,'' which occur when the annotation provided by an annotator is not something that you as a researcher would want assigned to your data. Have you observed mismatches in your data, and can you provide examples if so?
    \begin{itemize}
        \item Did you observe annotators making annotation errors? By errors, I mean that their annotations did not align with the stated annotation guidelines.
        \begin{itemize}
            \item If so, can you describe such a situation in detail?
            \item If so, why do you believe that those errors occurred? 
            \item If so, what did you do to identify, fix, and/or prevent those errors?
        \end{itemize}
        \item Did you experience situations where the correct annotation seemed ambiguous? By ambiguous, I mean that there could be multiple correct ways of annotating a particular piece of data, and the annotation instructions were under-specified.
        \begin{itemize}
            \item If so, can you describe such a situation in detail?
            \item  If so, why do you believe that the annotation was ambiguous? 
            \item If so, what did you do to resolve the ambiguity in that particular case? Were you able to do anything to prevent similar ambiguities from occurring in the future?
            \item Have you ever changed your annotation instructions to make them less ambiguous? If so, how?
        \end{itemize}
        \item Did you experience situations where the correct annotation seemed impossible? By impossible, I mean that there was not enough information available to know how to annotate it.
        \begin{itemize}
            \item If so, can you describe such a situation in detail?
            \item If so, why do you believe that the annotation was impossible? 
            \item If so, what did you do in that particular case? Were you able to do anything to prevent similar impossibilities from occurring in the future? 
            \item Have you ever excluded a piece of data from annotation because it seemed impossible to annotate?
        \end{itemize}
        \item Did you experience situations where the correct annotation seemed subjective? By subjective, I mean that the correct annotation depends on personal values, feelings, tastes, or opinions.
        \begin{itemize}
            \item If so, can you describe such a situation in detail?
            \item If so, why do you believe that the annotation was subjective? 
            \item If so, what did you do to manage that subjectivity? 
            \item Have you ever changed your annotation tasks to make them less subjective? If so, can you describe the change?
        \end{itemize}
        \item Did you experience situations where you felt as though whether an annotation was acceptable depended on the identity of the annotator? If so, can you describe such a situation in detail?
        \item When, if at all, do you think that mismatches become serious enough to affect data usability?
        \item Do mismatches matter on their own? Or do only the consequences of mismatches matter - in other words, if your data has annotation mismatches but still produces a high-quality predictive model, is this acceptable?
    \end{itemize}
\end{itemize}

\noindent \textbf{Going Forward:}
\begin{itemize}
    \item During your experience with the data annotation process, did you develop any best practices that you can share with us? 
    \item Is there anything that you tried but didn’t work? 
    \item Are there any challenges that you do not know how to address? 
\end{itemize}

\noindent \textbf{Closing:}
\begin{itemize}
    \item Is there anything else you’d like to share with us?
\end{itemize}

\end{document}